\UseRawInputEncoding
\documentclass[reprint,nofootinbib,showpacs]{revtex4}
\usepackage{graphicx}  
\usepackage{dcolumn}   
\usepackage{bm}        
\usepackage{amssymb}
\usepackage{multirow}
\usepackage{amsmath}
\usepackage{xcolor}
\usepackage{hyperref}
\usepackage[figurename=Figure]{caption}
\usepackage{caption}
\usepackage{subcaption}

\begin{document}

\title{One vanishing minor in  neutrino mass matrix using trimaximal mixing}
 \author{Iffat Ara Mazumder${}$}
 \email{iffat\_rs@phy.nits.ac.in}
 \author{Rupak~Dutta${}$}
 \email{rupak@phy.nits.ac.in}
 \affiliation{${}$ National Institute of Technology Silchar, Silchar 788010, India}

\begin{abstract}
We investigate the implications of one vanishing minor in neutrino mass matrix using trimaximal mixing matrix. In this context, we analyse all six patterns of one vanishing minor zero in neutrino mass matrix and present correlations of the neutrino oscillation parameters. All the six patterns are found to be phenomenologically viable with the present neutrino oscillation data. We also predict the values of effective Majorana mass, the effective electron anti-neutrino mass and the total neutrino mass for all the patterns. The value obtained for the effective neutrino mass is within the reach future neutrinoless double $\beta$ decay experiments. We also propose a flavor model where such patterns 
can be generated within the seesaw model. 
\end{abstract}

\maketitle
\section{Introduction}
Evidence of neutrino oscillations observed in multitude of experiments confirms that neutrinos mix with
each other and have non zero mass~\cite{Super-Kamiokande:1998kpq}. Neutrino oscillation phenomena can be parametrized in terms of six 
independent 
parameters, namely three mixing angles~$(\theta_{13}, \theta_{12}, \theta_{23})$, one Dirac CP violating phase~$(\delta_{\rm CP})$, and two mass squared differences~$(\Delta m_{21}^2\,, \Delta m_{31}^2)$. Although we have very precise values of the mixing angles and the absolute value of the mass squared differences, there are still some unknowns such as the octant of $\theta_{23}$, $\delta_{\rm CP}$ and the sign of $\Delta m_{31}^2$. There are two possible mass ordering of the neutrino mass spectrum: normal mass ordering (NO) $m_1 < m_2 << m_3$ and inverted mass ordering (IO) $m_3 << m_1 < m_2$ depending on the sign of $\Delta m_{31}^2$.

Neutrino oscillation experiments are sensitive only to the mass squared differences. They can not provide any
information regarding the absolute mass scale of neutrinos which is one of the most sought after questions in particle physics today.
Knowledge of the absolute mass scale of neutrinos is of great importance not only in particle physics but also in understanding the large 
scale 
structure of our universe. Neutrinos possess very tiny mass and unlike all other fermions in the Standard Model~(SM), they do not seem to get
their mass through Higgs mechanism. Hence, it may, in principle, help shape our understanding of the origin of particle mass which is still 
one of the most fundamental questions of particle physics. Unlike all other fermions in the SM, we observe only left handed neutrinos and
right handed anti-neutrinos. We have not found any right handed neutrino and left handed anti-neutrino so far in experiments. This brings us to
the next relevant question whether neutrino is a Dirac particle or a Majorana particle. Neutrino interactions could violate CP as well which
will be crucial in explaining the matter antimatter asymmetry in the universe. Moreover, there could be additional sterile neutrinos. 

There are several experimental efforts to find the absolute mass of the neutrino. The $\beta$ decay experiment performed at KATRIN can, in 
principle, measure the effective electron anti-neutrino mass by studying the end point region of the $\beta$ decay spectrum. This is 
completely model independent determination, i.e, it depends neither on any cosmological models nor on the nature of neutrinos. At present,
the improved upper bound on the effective electron anti-neutrino mass is reported to be $m_{\nu} < 0.8\,{\rm eV}$ at $90\%$ confidence level. 
The KATRIN experiment will continue to take data over the next several years and it is expected that the mass sensitivity will reach up to
$0.2\,{\rm eV}$. Future experiments like Project $8$~\cite{Project8:2022wqh}, designed to measure the absolute mass scale of the neutrino, hopes to reach a 
goal of $40\,{\rm meV/c^2}$ neutrino mass sensitivity.
Indirectly, one can have information on neutrino mass from cosmological observations. These cosmological observations are 
sensitive to the total neutrino mass and to the number of neutrino species. There are several
results related to the total neutrino mass coming from various cosmological observations. Most of these indirect methods put a limit
on the total neutrino mass to be less than $0.2\,{\rm eV}$. These results are, however, model dependent.
They rely heavily on several cosmological assumptions. Current upper bound on the total neutrino mass is reported by the Planck satellite  to be $\sum\,m_i < 0.12\,{\rm eV}$ at $95\%$ confidence level combining  BAO  data  with  CMB  data ~\cite{Zhang:2020mox}. If KATRIN's mass sensitivity reach up to $0.2\,{\rm eV}$ in future, it can put 
severe constraint on several cosmological models. Rare double $\beta$ decay process with two anti neutrinos in the final state is allowed in 
the SM. In general double beta decay processes are powerful probes of beyond the SM physics. More specifically, if one observes neutrinoless 
double $\beta$ decay in experiments, it would confirm that neutrinos are Majorana in nature. One can determine the effective Majorana 
mass~$M_{ee}$ by studying neutrinoless double beta decay. There exists several limits on the value of $M_{ee}$ using different isotopes.
At present, the best limits are reported to be $m_{ee} < (0.079 - 0.18)\,{\rm eV}$, $m_{ee} < (0.075 - 0.35)\,{\rm eV}$ and
$m_{ee} < (0.061 - 0.165)\,{\rm eV}$~\cite{GERDA:2013vls,CUORE:2015hsf,EXO-200:2014ofj,KamLAND-Zen:2012mmx}, respectively.  

There are several theoretical efforts in explaining the origin of neutrino mass. The most natural way to understand neutrino mass is through
seesaw mechanism. The neutrino mass matrix in the framework of Type-I seesaw mechanism is given by $M_{\nu}=-M_DM_R^{-1}M_D^{T}$,
where $M_D$ is the Dirac neutrino mass matrix and $M_R$ is the Majorana mass matrix of the right handed neutrinos.    
Phenomenology of Majorana neutrino mass matrix has been studied extensively assuming zero textures of the neutrino mass matrix which may be 
realised from the zeros in $M_D$ or $M_R$. In literature, there have been phenomenological studies with texture one-zero 
~\cite{Lashin:2011dn,Singh:2018tqu,Gautam:2018izb}, two-zeros~\cite{Frampton:2002yf,Xing:2002ta,Xing:2002ap,Lavoura:2004tu,Dev:2006qe,Kumar:2011vf,Fritzsch:2011qv,Ludl:2011vv,Meloni:2012sx,Grimus:2012zm,Dev:2014dla,Gautam:2016qyw,Channey:2018cfj,Singh:2019baq} and more within the context of 
Pontecorvo~Maki~Nakagawa~Sakata~(PMNS), tribimaximal~(TB) and trimaximal~(TM) mixing matrix. Similarly, in 
Ref.~\cite{Lashin:2007dm,Lashin:2009yd,Dev:2010if,Dev:2010pf,Tavartkiladze:2022pzf,Araki:2012ip}, the authors have studied the phenomenological implication of vanishing minors in the neutrino mass 
matrix. Moreover, in Ref.~\cite{Liao:2013saa,Dev:2013xca,Wang:2013woa,Whisnant:2015ovx,Dev:2015lya,Wang:2016tkm} and Ref.~\cite{Goswami:2008uv,Dev:2009he,Dev:2010pe,Liu:2013oxa,Dev:2013nua}, the authors have explored
the implication of cofactor zero and hybrid texture of the neutrino mass matrix. In case of zero textures, it is found that three or more 
zeros in neutrino mass matrix can not accommodate the current neutrino oscillation data. In Ref.~\cite{Frampton:2002yf}, the authors have 
found that out of fifteen possible two texture zeros cases only seven cases with PMNS mixing are allowed experimentally. Also out of fifteen 
possible two cofactors zero patterns only seven patterns are acceptable~\cite{Lashin:2007dm}. In case of TM mixing along with magic 
symmetry~\cite{Gautam:2016qyw}, the authors have found that only two cases are valid for two texture zero. TM mixing with one texture zero 
was studied in Ref.~\cite{Gautam:2018izb} and found that all six patterns are compatible with current neutrino oscillation data. For TB mixing 
along with the condition of texture zeros or vanishing minor~\cite{Dev:2010pf} only five patterns are allowed. In this work we study the 
implication of one vanishing minor in the neutrino mass matrix using trimaximal mixing.

Our paper is organized as follows. In Section.~\ref{section:2}, we briefly discuss the neutrino mass matrix using trimaximal mixing matrix.
We find all the mixing parameters such as $\theta_{13}$, $\theta_{23}$, $\theta_{12}$ and the CP violating parameter $\delta_{CP}$ in terms
of the unknown parameters $\theta$ and $\phi$ of the trimaximal mixing matrix. In Section.~\ref{section:3}, we describe the formalism of one 
vanishing minor in neutrino mass matrix and identify all the possible patterns of one vanishing minor. We provide all the detail numerical 
analysis and discussion of each pattern in Sections.~\ref{section:4} and ~\ref{section:5}. The fine-tuning of neutrino mass matrix is presented in Section.~\ref{section:6}. In Section.~\ref{section:7}, we present the symmetry realization and
conclude in Section.~\ref{section:8}.

\section{neutrino mass matrix}
\label{section:2}
The most widely studied lepton flavor mixing is TB  mixing pattern~\cite{Harrison:2002er,Harrison:2002kp,Xing:2002sw,Harrison:2003aw} introduced by 
Harrison, Perkins and Scott. TB mixing pattern provides remarkable agreement with the atmospheric and solar neutrino oscillation data.
The TB mixing pattern is given by
\begin{equation} 
\label{eq:1}
 U_{TB}=\begin{pmatrix}
  \sqrt{\frac{2}{3}}& \sqrt{\frac{1}{3}} & 0\\
  
 -\sqrt{\frac{1}{6}} & \sqrt{\frac{1}{3}} &\sqrt{\frac{1}{2}}\\
 
 -\sqrt{\frac{1}{6}} & \sqrt{\frac{1}{3}} & -\sqrt{\frac{1}{2}}
\end{pmatrix}.
\end{equation}
The TB mixing matrix possesses two types of symmetries: $\mu - \tau$ symmetry and magic symmetry.
Although TB mixing matrix correctly predicted the value of atmospheric mixing angle $\theta_{23}$ and the solar mixing angle
$\theta_{12}$, it, however, failed to explain a non zero value of the reactor mixing angle $\theta_{13}$ that was experimentally confirmed
by T2K~\cite{T2K:2011ypd}, MINOS~\cite{MINOS:2011amj}, Double Chooz~\cite{DoubleChooz:2011ymz}, Daya Bay~\cite{DayaBay:2012fng} and 
RENO~\cite{RENO:2012mkc} experiments. The possibility of an exact $\mu - \tau$ symmetry in the mass matrix was completely ruled out by a 
relatively large value of $\theta_{13}$. 
Modifications in the TB mixing pattern~\cite{Kumar:2010qz,He:2011gb,Grimus:2008tt} was made to accommodate the present data.
The TM mixing matrix was constructed by multiplying the TB mixing matrix by an unitary matrix and can be written as
\begin{equation} 
\label{eq:2}
  U_{TM_1}=\begin{pmatrix}
 \sqrt{\frac{2}{3}} & \frac{1}{\sqrt{3}}\cos\theta & \frac{1}{\sqrt{3}}\sin\theta\\
 -\frac{1}{\sqrt{6}}& \frac{\cos\theta}{\sqrt{3}}-\frac{e^{i\phi}\sin\theta}{\sqrt{2}}&\frac{\sin\theta}{\sqrt{3}}+\frac{e^{i\phi}\cos\theta}{\sqrt{2}}\\
 -\frac{1}{\sqrt{6}}& \frac{\cos\theta}{\sqrt{3}}+\frac{e^{i\phi}\sin\theta}{\sqrt{2}}&\frac{\sin\theta}{\sqrt{3}}-\frac{e^{i\phi}\cos\theta}{\sqrt{2}}\,
\end{pmatrix}. 
 \end{equation}
and
\begin{equation} 
\label{eq:3}
  U_{TM_2}=\begin{pmatrix}
 \sqrt{\frac{2}{3}}\cos\theta & \frac{1}{\sqrt{3}} & \sqrt{\frac{2}{3}}\sin\theta\\
 -\frac{\cos\theta}{\sqrt{6}}+\frac{e^{-i\phi}\sin\theta}{\sqrt{2}}& \frac{1}{\sqrt{3}}&-\frac{\sin\theta}{\sqrt{6}}-\frac{e^{-i\phi}\cos\theta}{\sqrt{2}}\\
 -\frac{\cos\theta}{\sqrt{6}}-\frac{e^{-i\phi}\sin\theta}{\sqrt{2}}& \frac{1}{\sqrt{3}}&-\frac{\sin\theta}{\sqrt{6}}+\frac{e^{-i\phi}\cos\theta}{\sqrt{2}}\,
\end{pmatrix}. 
 \end{equation}
where $\theta$ and $\phi$ are two free parameters. The neutrino mass matrix corresponding to TM mixing matrix can be written as
\begin{equation} 
\label{eq:4}
 M_{\rho\sigma}=(VM_{diag}V^{T})_{\rho\sigma}\,\, {\rm with}\,\, \rho\,, \sigma = e\,,\mu\,,\tau\,,
\end{equation}
where $M_{diag} = {\rm diag}(m_1, m_2, m_3)$ is the diagonal matrix containing three mass state, $V=U_{TM}P$ and $P$ is the phase matrix 
written as
\begin{equation} 
\label{eq:5}
 P=\begin{pmatrix}
  1& 0 & 0\\
 0 & e^{i\alpha} &0\\
 0 & 0 & e^{i\beta}\
\end{pmatrix}\,.
\end{equation}
Here $\alpha$ and $\beta$ are the two CP violating Majorana phases.

\subsection{TM$_1$ Mixing matrix}
With TM$_1$ mixing matrix, the elements of neutrino mass matrix can be written as  
\begin{eqnarray} 
\label{eq:6}
  &&M_{ee}=\frac{2}{3}\,m_1+\frac{1}{3}\cos^2\theta\,m_2\,e^{2i\alpha}+\frac{1}{3}\sin^2\theta\,m_3\,e^{2\,i\,\beta},  
\nonumber \\ 
  &&M_{ e\mu}=(-\frac{1}{3})m_1+(\frac{1}{3}\cos^2\theta-\frac{1}{\sqrt{6}}\sin\theta \cos\theta e^{i\phi})m_2 e^{2i\alpha}+
(\frac{1}{3}\sin^2\theta+\frac{1}{\sqrt{6}}\sin\theta \cos\theta e^{i\phi}) m_3 e^{2i\beta},  \nonumber \\
  &&M_{ e\tau}=(-\frac{1}{3})m_1+(\frac{1}{3}\cos^2\theta+\frac{1}{\sqrt{6}}\sin\theta \cos\theta e^{i\phi})m_2 e^{2i\alpha}+(\frac{1}{3}\sin^2\theta-\frac{1}{\sqrt{6}}\sin\theta \cos\theta e^{i\phi}) m_3 e^{2i\beta},  \nonumber \\
  &&M_{ \mu\mu}=\frac{1}{6}m_1+(\frac{1}{\sqrt{3}}\cos\theta-\frac{1}{\sqrt{2}}\sin\theta  e^{i\phi})^2m_2 e^{2i\alpha}+
(\frac{1}{\sqrt{3}}\sin\theta+\frac{1}{\sqrt{2}}\cos\theta e^{i\phi})^2 m_3 e^{2i\beta},  \nonumber \\
  &&M_{ \mu\tau}=\frac{1}{6}m_1+(\frac{1}{3}\cos^2\theta-\frac{1}{2}\sin^2\theta e^{2i\phi}) m_2 e^{2i\alpha}+(\frac{1}{3}
\sin^2\theta-\frac{1}{2}\cos^2\theta e^{2i\phi}) m_3 e^{2i\beta},  \nonumber \\
  &&M_{ \tau\tau}=\frac{1}{6} m_1+(\frac{1}{\sqrt{3}}\cos\theta+\frac{1}{\sqrt{2}}\sin\theta e^{i\phi})^2 e^{2i\alpha}+
(\frac{1}{\sqrt{3}}\sin\theta-\frac{1}{\sqrt{2}}\cos\theta e^{i\phi})^2 m_3 e^{2i\beta}.  
 \end{eqnarray}
The three neutrino mixing angles $\theta_{12}$, $\theta_{23}$ and $\theta_{13}$ can be expressed in terms of $\theta$ and $\phi$, the free 
parameters of the TM$_1$ matrix, as 
\begin{eqnarray} 
\label{eq:7}
&& s_{12}^2=\frac{|(U_{12})_{TM_1}|^2}{1-|(U_{13})_{TM_1}|^2} = 1-\frac{2}{3-\sin^2\theta}\,, \nonumber \\
&& s_{23}^2=\frac{|(U_{23})_{TM_1}|^2}{1-|(U_{13})_{TM_1}|^2} = \frac{1}{2}\Big(1+\frac{\sqrt{6}\sin2\theta \cos\phi}{3-\sin^2\theta}\Big)\,,
\nonumber \\
&& s_{13}^2=|(U_{13})_{TM_1}|^2 = \frac{1}{3}\sin^2\theta\,,
\end{eqnarray}
where $s_{ij}=\sin\theta_{ij}$ and $c_{ij}=\cos\theta_{ij}$ for $i,j=1,2,3$.
Using the standard parametrization of the PMNS matrix, the Jarlskog invariant, a measure of CP violation, is defined as~\cite{Jarlskog:1985ht}
\begin{eqnarray}
\label{eq:8} 
 J&=& s_{12}s_{13}s_{23}c_{12}c_{13}^2c_{23}\sin\delta.
 \end{eqnarray}
Again, using the elements from TM$_1$ mixing matrix, the Jarlskog invariant can be expressed as
\begin{equation}
\label{eq:9}
 J =\frac{1}{6\sqrt{6}}\sin2\theta \sin\phi. 
\end{equation}
Combining Eq.~\ref{eq:8} and Eq.~\ref{eq:9}, we can write $\delta$ in terms of $\theta$ and $\phi$ as 
\begin{equation}
\label{eq:10}
 \csc^2\delta=\csc^2\phi-\frac{6\sin^{2}2\theta \cot^2\phi}{(3-\sin^2\theta)^2}.
\end{equation}
The nature of neutrino can be determined from the effective Majorana mass term. It also measures the rate of neutrinoless double beta decay. 
The effective Majorana mass $|M_{ee}|$ for the TM$_1$ mixing matrix can be written as
\begin{equation} 
\label{eq:11}
|M_{ee}|=\Big|\frac{1}{3}(2m_1+ m_2\cos^2\theta  e^{2i\alpha} + m_3\sin^2\theta  e^{2i\beta})\Big|.
\end{equation}
Similarly, the effective electron anti-neutrino mass can be expressed as 
\begin{equation} 
\label{eq:12}
 M_{\nu}^2=\sum\limits_{i=1}^{3}U_{ie}^2=\frac{1}{3}(2m_1^2+ m_2^2\cos^2\theta + m_3^2\sin^2\theta).
\end{equation}
\subsection{TM$_2$ Mixing matrix}
Using TM$_2$ mixing matrix, we can write the elements of neutrino mass matrix as   
   \begin{eqnarray} 
\label{eq:13}
 &&M_{ee}=(\frac{2}{3}\cos^2\theta)\,m_1+\frac{1}{3}\,m_2\,e^{2i\alpha}+(\frac{2}{3}\sin^2\theta)\,m_3\,e^{2\,i\,\beta},  
\nonumber \\ 
  &&M_{ e\mu}=(-\frac{1}{3}\cos^2\theta+\frac{1}{\sqrt{3}}\sin\theta \cos\theta e^{-i\phi}) m_1+\frac{1}{3}m_2 e^{2i\alpha}+
(-\frac{1}{3}\sin^2\theta-\frac{1}{\sqrt{3}}\sin\theta \cos\theta e^{-i\phi}) m_3 e^{2i\beta},  \nonumber \\
  &&M_{ e\tau}=(-\frac{1}{3}\cos^2\theta-\frac{1}{\sqrt{3}}\sin\theta \cos\theta e^{-i\phi}) m_1+\frac{1}{3}m_2 e^{2i\alpha}+
(-\frac{1}{3}\sin^2\theta+\frac{1}{\sqrt{3}}\sin\theta \cos\theta e^{-i\phi}) m_3 e^{2i\beta},  \nonumber \\
  &&M_{ \mu\mu}=(-\frac{1}{\sqrt{6}}\cos\theta+\frac{1}{\sqrt{2}}\sin\theta  e^{-i\phi})^2 m_1+\frac{1}{3}m_2 e^{2i\alpha}+
(\frac{1}{\sqrt{6}}\sin\theta+\frac{1}{\sqrt{2}}\cos\theta e^{-i\phi})^2 m_3 e^{2i\beta},  \nonumber \\
  &&M_{ \mu\tau}=(\frac{1}{6}\cos^2\theta-\frac{1}{2}\sin^2\theta e^{-2i\phi}) m_1+\frac{1}{3}m_2 e^{2i\alpha}+(\frac{1}{6}
\sin^2\theta-\frac{1}{2}\cos^2\theta e^{-2i\phi}) m_3 e^{2i\beta},  \nonumber \\
  &&M_{ \tau\tau}=(\frac{1}{\sqrt{6}}\cos\theta+\frac{1}{\sqrt{2}}\sin\theta  e^{-i\phi})^2 m_1+\frac{1}{3}m_2 e^{2i\alpha}+
(-\frac{1}{\sqrt{6}}\sin\theta+\frac{1}{\sqrt{2}}\cos\theta e^{-i\phi})^2 m_3 e^{2i\beta}.  
 \end{eqnarray}
The three neutrino mixing angles $\theta_{12}$, $\theta_{23}$ and $\theta_{13}$ can be expressed as 
\begin{eqnarray}
\label{eq:14}
&& s_{12}^2= \frac{1}{3-2\sin^2\theta}\,, 
\nonumber \\
&& s_{23}^2 = \frac{1}{2}\Big(1+\frac{\sqrt{3}\sin2\theta \cos\phi}{3-2\sin^2\theta}\Big)\,,
\nonumber \\
&& s_{13}^2 = \frac{2}{3}\sin^2\theta\,.
\end{eqnarray}
Again, using the elements from TM$_2$ mixing matrix, the Jarlskog invariant can be expressed as
\begin{equation}
\label{eq:15}
 J =\frac{1}{6\sqrt{3}}\sin2\theta \sin\phi.
\end{equation}
We can express the Dirac CP violating parameter $\delta_{CP}$ in terms of $\theta$ and $\phi$ as
\begin{equation} 
\label{eq:16}
 \csc^2\delta=\csc^2\phi-\frac{3\sin^{2}2\theta \cot^2\phi}{(3-2\sin^2\theta)^2}.
\end{equation}
The effective Majorana mass $|M_{ee}|$ for the TM$_2$ mixing matrix can be written as
 \begin{equation}
\label{eq:17}
  |M_{ee}|=\Big|\frac{1}{3}(2m_1\cos^2\theta + m_2 e^{2i\alpha} + 2m_3\sin^2\theta  e^{2i\beta})\Big|.
 \end{equation}
The effective electron anti-neutrino mass can be expressed as 
 \begin{equation} 
\label{eq:18}
  M_{\nu}^2=\frac{1}{3}(2m_1^2+ m_2^2\cos^2\theta + 2m_3^2\sin^2\theta).
 \end{equation}
\section{One vanishing minor in neutrino mass matrix}\label{section:3}
There are six independent minors corresponding to six independent elements in the neutrino mass matrix. We denote the minor corresponding to 
$ij^{th}$ element of $M_{ij}$ as $C_{ij}$. 
The six possible patterns of one minor zero in neutrino mass matrix are listed in Table.~\ref{tab:1}.
\begin{table}[h!]
\begin{tabular}{|c|c|}
 \hline
Pattern & Constraining equation\\
\hline
I& $ C_{33}=0$\\
\hline
II & $ C_{22}=0$\\
\hline
III & $ C_{31}=0$\\
\hline
IV & $ C_{21}=0$\\
\hline
V & $ C_{32}=0$\\
\hline
VI & $ C_{11}=0$\\
\hline
\hline
\end{tabular}
\caption{ one minor zero patterns. }
\label{tab:1}
\end{table}
The condition for one vanishing minor can be written as
\begin{equation} 
\label{eq:19}
M_{a\,b}M_{c\,d}-M_{u\,v}M_{w\,x}=0\,.
\end{equation}
More specifically, we can write Eq.~\ref{eq:19} in terms of a complex equation as 
\begin{equation} 
\label{eq:20}
 m_{1}m_{2}X_{3}e^{2i\alpha}+m_{2}m_{3}X_{1}e^{2i(\alpha+\beta)}+m_{3}m_{1}X_{2}e^{2i\beta}=0\,,
\end{equation}
where 
\begin{equation}
 X_{k}=(U_{ai}U_{bi}U_{cj}U_{dj}-U_{ui}U_{vi}U_{wj}U_{xj})+(i\leftrightarrow j)\,,
\end{equation}
with $(i,j, k)$ as the cyclic permutation of $(1,2,3)$.
Using Eq.~\ref{eq:20}, one can write the two mass ratios as
\begin{eqnarray} 
 \label{eq:21}
&& \frac{m_1}{m_2}=\frac{\Re(X_{3}e^{2i\alpha})\Im(X_1e^{2i(\alpha+\beta)})-\Re(X_1e^{2i(\alpha+\beta)})\Im(X_{3}e^{2i\alpha})}
{\Re(X_{2}e^{2i\beta})\Im(X_{3}e^{2i\alpha})-\Re(X_{3}e^{2i\alpha})\Im(X_{2}e^{2i\beta})}\,, \nonumber \\
&& \frac{m_3}{m_2}=\frac{\Re(X_{3}e^{2i\alpha})\Im(X_1e^{2i(\alpha+\beta)})-\Re(X_1e^{2i(\alpha+\beta)})\Im(X_{3}e^{2i\alpha})}
{\Re(X_1e^{2i(\alpha+\beta)})\Im(X_{2}e^{2i\beta})-\Re(X_{2}e^{2i\beta})\Im(X_1e^{2i(\alpha+\beta)})}.
\end{eqnarray}
The value of $m_1$, $m_2$ and $m_3$ can be calculated using Eq.~\ref{eq:21} and mass square difference $\Delta m_{21}^2$. 
That is
\begin{eqnarray}
\label{eq:22}
&& m_{1}=\sqrt{\Delta m_{21}^2}\sqrt{\frac{(\frac{m_{1}}{m_{2}})^2} {|1-(\frac{m_{1}}{m_{2}})^2|}}, \nonumber \\ 
&& m_{2}=\sqrt{\Delta m_{21}^2}\sqrt{\frac{1} {|1-(\frac{m_{1}}{m_{2}})^2|}}\,, \nonumber \\
&& m_{3}=\sqrt{\Delta m_{21}^2}\sqrt{\frac{(\frac{m_{3}}{m_{2}})^2} {|1-(\frac{m_{1}}{m_{2}})^2|}}\,.
 \end{eqnarray}
Similarly, the ratio of squared mass difference is defined as
 \begin{equation} 
\label{eq:26}
  r\equiv \Big|\frac{\Delta m_{21}^2}{\Delta m_{32}^2}\Big|=\Big|\frac{1-(\frac{m_{1}}{m_{2}})^2}{(\frac{m_{3}}{m_{2}})^2-1}\Big|\,,
 \end{equation}
where $\Delta m_{21}^2$ and $\Delta m_{32}^2$ represent solar and atmospheric mass squared difference, respectively. Value of 
$r = (2.950\pm0.08)\times10^{-2}$ is determined by using the measured values of $\Delta m_{21}^2$ and $\Delta m_{32}^2$ reported in 
Ref.~\cite{Esteban:2020cvm}. 

\section{Results and discussion}
\label{section:4}
For our numerical analysis, we use the measured values of the oscillation parameters reported in Ref.~\cite{Esteban:2020cvm}. For completeness, we report them in Table.~\ref{tab:2}.
\begin{table}[htbp]
\begin{tabular}{|c|c|c|c|}
 \hline
 \hline
parameter & Normal ordering(best fit) & inverted ordering ($\Delta \chi^2=7.1)$\\
&bfp$\pm 1\sigma$ \hspace{1.5cm} $3\sigma$ ranges&bfp$\pm 1\sigma$ \hspace{1.5cm} $3\sigma$ ranges\\
\hline
$\theta_{12}^\circ
$&$33.44^{+0.77}_{-0.74}$ \hspace{1.5cm} 31.27$\rightarrow $ 35.86&$33.45^{+0.77}_{-0.74}$\hspace{1.5cm} 31.27$\rightarrow $ 35.87\\
\hline
$\theta_{23}^\circ$&$49.2^{+1.0}_{-1.3}$ \hspace{1.5cm} 39.5$\rightarrow $ 52.0&$49.5^{+1.0}_{-1.2}$\hspace{1.5cm} 39.8$\rightarrow $ 52.1\\
\hline
$\theta_{13}^\circ$&$8.57^{+0.13}_{-0.12}$ \hspace{1.5cm} 8.20$\rightarrow $ 8.97&$8.60^{+0.12}_{-0.12}$ \hspace{1.5cm} 8.24$\rightarrow $ 8.98\\
\hline
$\delta^\circ$&$194^{+52}_{-25}$ \hspace{1.5cm} 105$\rightarrow $ 405&$287^{+27}_{-32}$ \hspace{1.5cm} 192$\rightarrow $ 361\\
\hline
$\frac{\Delta m^2_{21}}{10^{-5}eV^2}$&$7.42^{+0.21}_{-0.20}$ \hspace{1.5cm} 6.82$\rightarrow $ 8.04&$7.42^{+0.21}_{-0.20}$ \hspace{1.5cm} 6.82$\rightarrow $ 8.04\\
\hline
$\frac{\Delta m^2_{3l}}{10^{-3}eV^2}$&$+2.515^{+0.028}_{-0.028}$ \hspace{1.5cm} +2.431$\rightarrow $ +2.599&$-2.498^{+0.028}_{-0.029}$ \hspace{1.5cm} -2.584$\rightarrow $ -2.413\\
\hline
\hline
\end{tabular}
\caption{neutrino oscillation parameters from NuFIT \cite{Esteban:2020cvm}. }
\label{tab:2}
\end{table}
We wish to find the value of the unknown parameters $\theta$ and $\phi$. It is evident from Eq.~\ref{eq:7} and Eq.~\ref{eq:14} that the neutrino oscillation parameters $\theta_{12}$ and $\theta_{13}$ depend only on $\theta$. To find the best fit value of $\theta$, we perform a naive $\chi^2$ analysis. The relevant $\chi^2$ is defined as
 \begin{equation} 
\label{eq:27}
  \chi^2(\theta)=\sum\limits_{i=1}^{2} \frac{\Big(\theta_{i}^{cal}- \theta_{i}^{exp}\Big)^2}{(\sigma_{i}^{exp})^2}\,,
 \end{equation}
where $\theta_i=(\theta_{12},\theta_{13})$. Here $\theta_{i}^{cal}$ represents the theoretical value of $\theta_{i}$ and $\theta_{i}^{exp}$ 
represents measured central value of $\theta_{i}$. The corresponding uncertainties in the measured value of $\theta_{i}$ is represented by 
$\sigma_{i}^{exp}$.

\begin{figure}[htbp]
\begin{subfigure}{0.35\textwidth}
\includegraphics[width=\textwidth]{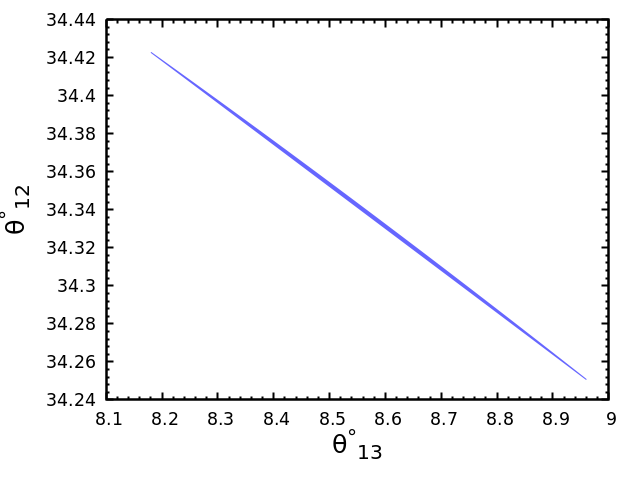}
\caption{}
\label{fig:TM1ff1}
\end{subfigure}
\begin{subfigure}{0.35\textwidth}
\includegraphics[width=\textwidth]{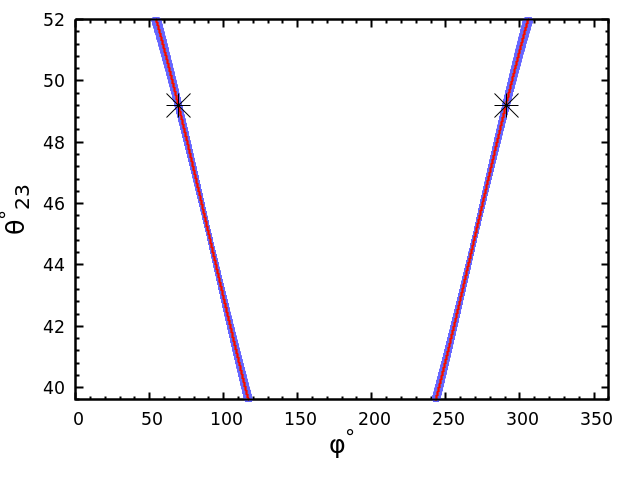}
\caption{}
\label{fig:TM1ff2}
\end{subfigure}
\begin{subfigure}{0.35\textwidth}
\includegraphics[width=\textwidth]{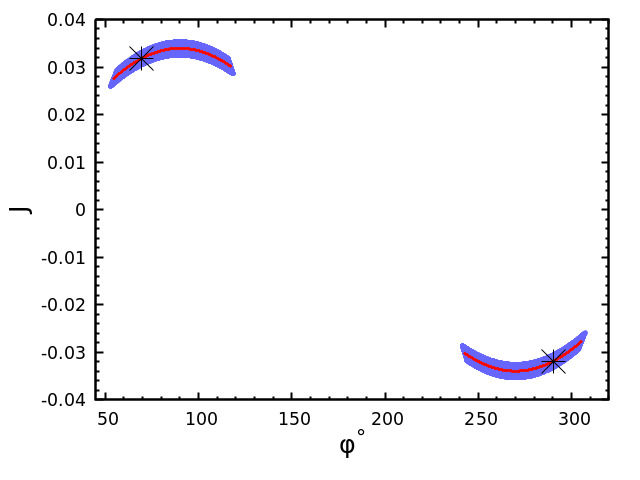}
\caption{}
\label{fig:TM1ff3}
\end{subfigure}
\begin{subfigure}{0.35\textwidth}
\includegraphics[width=\textwidth]{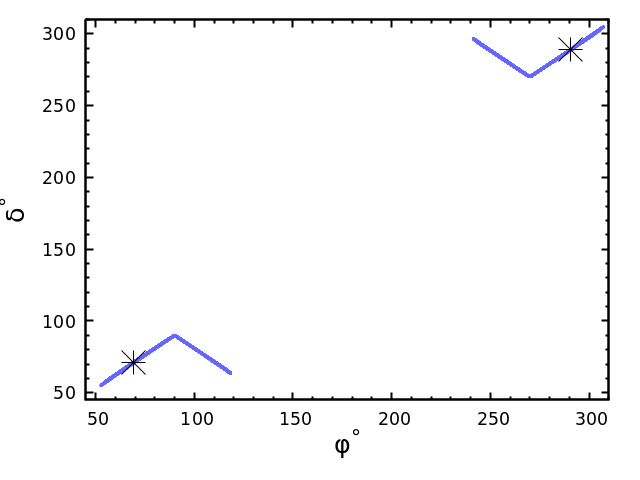}
\caption{}
\label{fig:TM1ff4}
\end{subfigure}
\caption{Various correlation plots for TM$_1$ mixing matrix.}
\end{figure}

For the TM$_1$ mixing matrix, the best fit value of $\theta$ is obtained to be $14.96^\circ$. The corresponding best fit values of 
$\theta_{12}$ and $\theta_{13}$ are $34.33^{\circ}$ and $8.57^{\circ}$, respectively. The $3\sigma$ allowed range of $\theta$ is found to be 
$(14.26^\circ - 15.64^\circ)$. Using the allowed range of $\theta$, we obtain the allowed ranges of $\theta_{12}$ and $\theta_{23}$ to be 
$(34.25^\circ - 34.42^\circ)$ and $(32.11^\circ - 57.88^\circ)$, respectively. We show in Fig.~\ref{fig:TM1ff1} the correlation of 
$\theta_{13}$ and $\theta_{12}$ for the TM$_1$ mixing matrix. 
To see the variation of $\theta_{23}$ with $\phi$, we use the allowed range of $\theta$ and vary $\phi$ within its full range
from $0^\circ$ to $360^\circ$. We show in Fig.~\ref{fig:TM1ff2} the variation of $\theta_{23}$ as a function of the unknown parameter $\phi$.  
We also obtain the best fit value of $\phi$ by using the measured best fit value of $\theta_{23}$. The best fit value is shown with a '$*$' 
mark in Fig.~\ref{fig:TM1ff2}. The best fit values of $\phi$ corresponding 
to the best fit value of $\theta_{23} = 49.2^{\circ}$ are $69.43^\circ$ and $290.57^\circ$, respectively. We show the variation of $J$ and 
$\delta$ as a function of $\phi$ in Fig.~\ref{fig:TM1ff3} and Fig.~\ref{fig:TM1ff4}, respectively. It is observed that the Jarlskog rephasing 
invariant $J$ and the Dirac CP violating phase $\delta$ are restricted to two regions. The corresponding best fit values of $J$ and $\delta$ 
are $[-3.184\times 10^{-2},3.185\times 10^{-2}]$ and $[71.11^\circ, 288.98^\circ]$, respectively. We also obtain the $3\sigma$ allowed ranges 
of $J$ and $\delta$ to be $[0,\pm 3.53\times 10^{-2}]$ and $[(55.05, 124.95)^\circ,\,(235.05, 304.95)^\circ]$, respectively.

\begin{figure}[htbp]
\begin{subfigure}{0.35\textwidth}
\includegraphics[width=\textwidth]{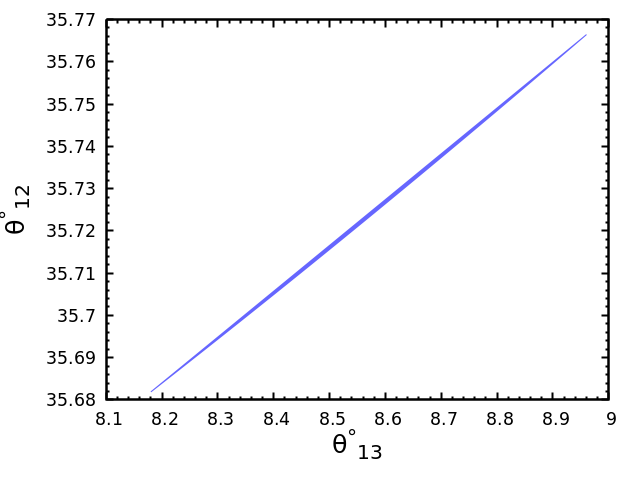}
\caption{}
\label{fig:TM2ff1}
\end{subfigure}
\begin{subfigure}{0.35\textwidth}
\includegraphics[width=\textwidth]{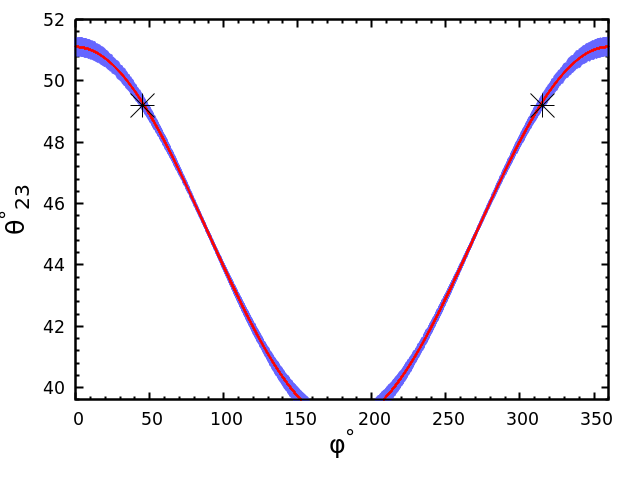}
\caption{}
\label{fig:TM2ff2}
\end{subfigure}
\begin{subfigure}{0.35\textwidth}
\includegraphics[width=\textwidth]{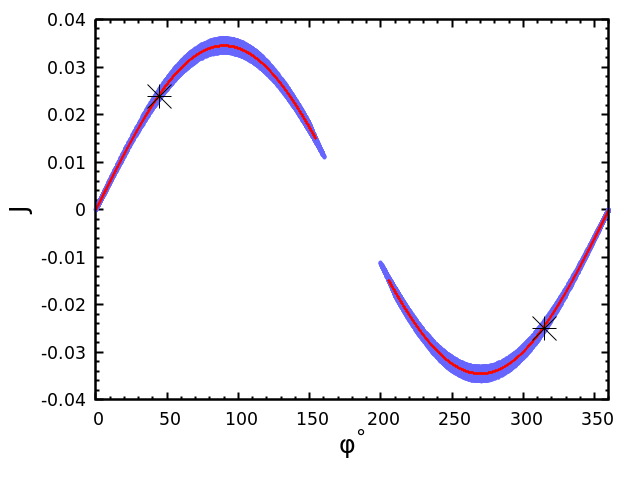}
\caption{}
\label{fig:TM2ff3}
\end{subfigure}
\begin{subfigure}{0.35\textwidth}
\includegraphics[width=\textwidth]{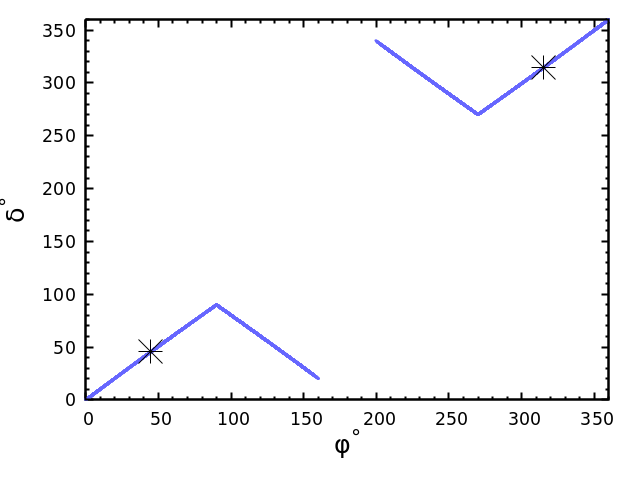}
\caption{}
\label{fig:TM2ff4}
\end{subfigure}
\caption{Various correlation plots for TM$_2$ mixing matrix. }
\end{figure}

For TM$_2$ mixing matrix, the best fit value of $\theta$ is obtained to be $10.50^\circ$. The corresponding best fit values of $\theta_{12}$ 
and $\theta_{13}$ are $35.72^{\circ}$ and $8.56^{\circ}$, respectively. The $3\sigma$ allowed range of $\theta$ is found to be 
$(10.03^\circ - 10.99^\circ)$. Using the allowed range of $\theta$, we obtain the allowed ranges of $\theta_{12}$ and $\theta_{23}$ to be 
$(35.68^\circ - 35.76^\circ)$ and $(39.50^\circ - 51.40^\circ)$, respectively. It should be noted that although the allowed range of 
$\theta_{12}$ obtained with TM$_2$ mixing matrix is consistent with the $3\sigma$ experimental range, the best fit value obtained for 
$\theta_{12}$, however, deviates from the experimental best fit value at more than $2\sigma$ significance. This is quite a generic feature of 
TM$_2$ mixing matrix because, by default, value of $\theta_{12}$ will be greater than or equal to the value obtained in case of TB mixing 
matrix. We show in Fig.~\ref{fig:TM2ff1} the correlation of $\theta_{13}$ and $\theta_{12}$ for the TM$_2$ mixing matrix.

To see the variation of $\theta_{23}$ with $\phi$, we use the allowed range of $\theta$ and vary $\phi$ within its full range
from $0^\circ$ to $360^\circ$. We show in Fig.~\ref{fig:TM2ff2} the variation of $\theta_{23}$ as a function of the unknown parameter $\phi$. The best fit value 
is shown with a '$*$' mark in Fig.~\ref{fig:TM2ff2}. We obtain the best fit value of $\phi$ by using the measured best fit value of $\theta_{23}$. The best fit values of $\phi$ corresponding to the best fit value of $\theta_{23} = 49.2^{\circ}$ are $44.86^\circ$ and $315.19^\circ$, respectively. 
We get two best fit values of $\phi$ because $\theta_{23}$ is invariant under the transformation $\phi \to (2\pi - \phi)$ which is evident from Eq.~\ref{eq:7} and Eq.~\ref{eq:14}. We also show the variation of $J$ and $\delta$ as a function of $\phi$ in Fig.~\ref{fig:TM2ff3} and Fig.~\ref{fig:TM2ff4}, respectively. It is observed that the Jarlskog rephasing invariant $J$ and the Dirac CP violating phase $\delta$ are restricted to two regions. The corresponding best fit values of $J$ and $\delta$ are $[2.37\times 10^{-2}, -2.50\times 10^{-2}]$ and $[45.48^\circ, 314.44^\circ]$, 
respectively. We also obtain the $3\sigma$ allowed ranges 
of $J$ and $\delta$ to be $[0,\pm 3.60\times 10^{-2}]$ and $[(0, 90)^\circ,\,(270, 360)^\circ]$, respectively.

In case of inverted mass ordering the $3\sigma$ allowed range of $\theta$ are found to be $(14.37^\circ - 15.64^\circ)$ and $(10.10^\circ - 10.99^\circ)$ for both  TM$_1$ and TM$_2$ mixing matrix, respectively. Using the $3\sigma$ allowed range of $\theta$, we obtain the $3\sigma$ allowed ranges of $\theta_{12}$ to be 
$(34.25^\circ - 34.41^\circ)$ and $(35.68^\circ - 35.76^\circ)$, respectively for both TM$_1$ and TM$_2$ mixing matrix. The $3\sigma$ allowed ranges of $\theta_{23}$ to be $(32.12^\circ - 57.87^\circ)$ and $(38.60^\circ - 51.39^\circ)$, respectively for both TM$_1$ and TM$_2$ mixing matrix. It is to be noted that the mixing angles are almost similar for both the normal and inverted mass ordering. So for our later discussion we will use the values of mixing angles for the normal mass ordering reported in Ref.~\cite{Esteban:2020cvm}. 

\section{Phenomenology of one vanishing minor}
\label{section:5}
We wish to investigate the phenomenological implication of one vanishing minor in the neutrino mass matrix on the total neutrino mass, the effective Majorana mass term and the electron anti-neutrino mass. It is evident from Eq.~\ref{eq:22} that neutrino mass $m_{i}$ depends on $\theta$, $\phi$, $\alpha$, $\beta$ and
the mass squared difference $\Delta m_{21}^2$. We use the best fit value and the $3\sigma$ allowed range of $\theta$ and $\phi$ of 
section.~\ref{section:4} that are determined by the measured values of the mixing angles $\theta_{13}$, $\theta_{12}$ and $\theta_{23}$. The two
unknown Majorana phases $\alpha$ and $\beta$ are varied within their full range from $0^{\circ}$ to $360^{\circ}$. Moreover, we use the
$3\sigma$ allowed ranges of $\Delta m_{21}^2$ and $r$ to constrain the values of the neutrino masses. Now we proceed to analyse all the six 
patterns of one vanishing minor one by one.

\subsection{\textbf{Pattern I: $\boldsymbol {C_{33}=0}$}} 
\begin{figure}[htbp]
\begin{subfigure}{0.32\textwidth}
\includegraphics[width=\textwidth]{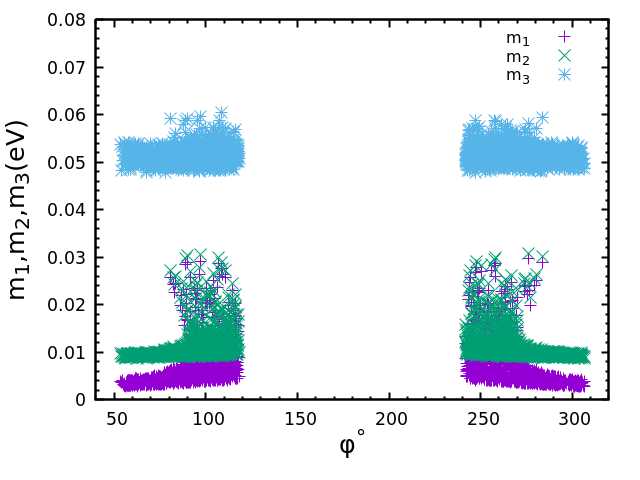}
\caption{}
\label{fig:3a}
\end{subfigure}
\begin{subfigure}{0.32\textwidth}
\includegraphics[width=\textwidth]{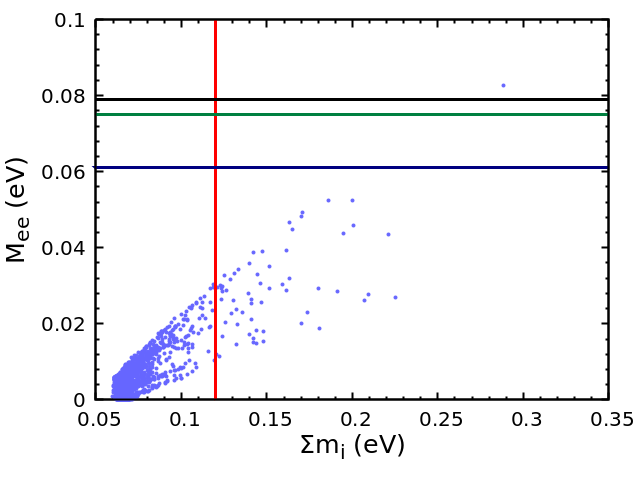}
\caption{}
\label{fig:3b}
\end{subfigure}
\begin{subfigure}{0.32\textwidth}
\includegraphics[width=\textwidth]{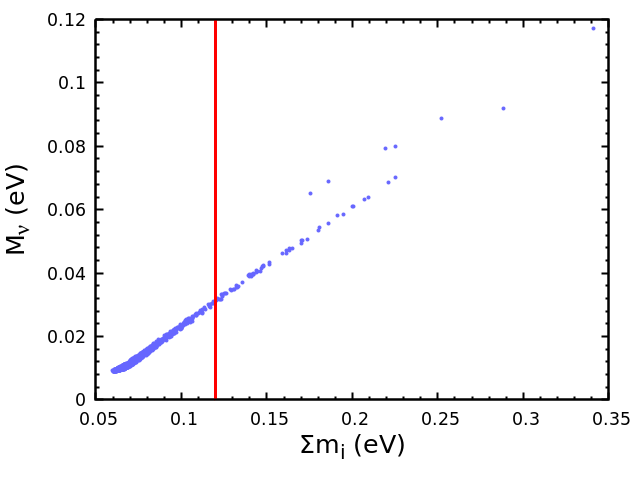}
\caption{}
\label{fig:3c}
\end{subfigure}
\label{fig:3}
\captionsetup{justification=raggedright,singlelinecheck=false}
\caption{
Various correlation plots for C$_{33}$ pattern using TM$_1$ mixing matrix. The vertical red line is the upper bound of the total neutrino mass reported in Ref.~\cite{Zhang:2020mox}. The black, green and blue lines are the experimental upper bounds of the effective Majorana mass reported in Ref.~\cite{GERDA:2013vls,CUORE:2015hsf,EXO-200:2014ofj,KamLAND-Zen:2012mmx}.}
\end{figure}

\begin{figure}[htbp]
\begin{subfigure}{0.32\textwidth}
\includegraphics[width=\textwidth]{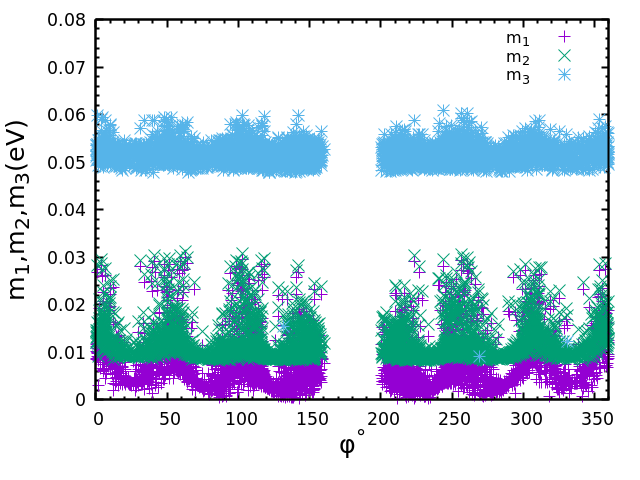}
\caption{}
\label{fig:4a}
\end{subfigure}
\hfill
\begin{subfigure}{0.32\textwidth}
\includegraphics[width=\textwidth]{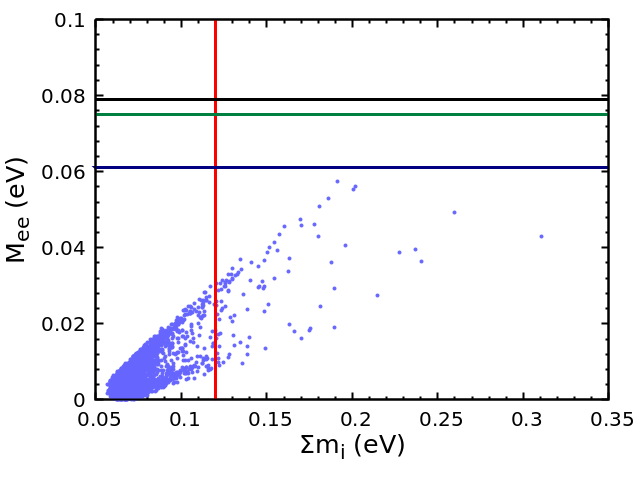}
\caption{}
\label{fig:4b}
\end{subfigure}
\hfill
\begin{subfigure}{0.32\textwidth}
\includegraphics[width=\textwidth]{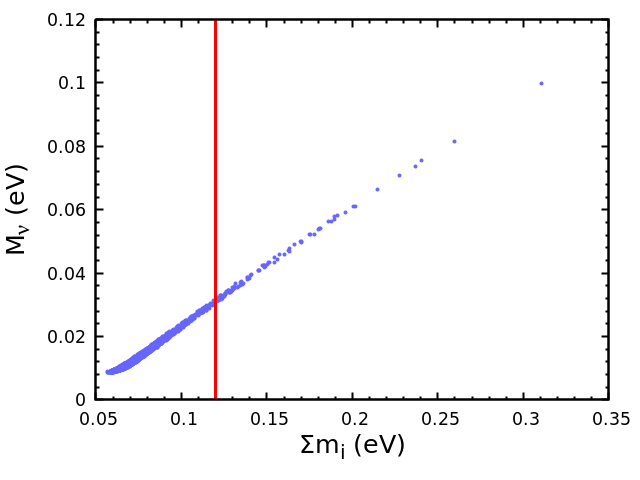}
\caption{}
\label{fig:4c}
\end{subfigure}
\captionsetup{justification=raggedright,singlelinecheck=false}
\caption{Various correlation plots for C$_{33}$ pattern using TM$_2$ mixing matrix. The vertical red line represents the upper bound of $\sum m_i$ reported in Ref.~\cite{Zhang:2020mox}. The black, green and blue lines are the experimental upper bounds of $M_{ee}$ obtained from Refs.~\cite{GERDA:2013vls,CUORE:2015hsf,EXO-200:2014ofj,KamLAND-Zen:2012mmx}.}
\label{fig:4}
\end{figure}

let us first consider minor zero for the $(3,3)$ element of the neutrino mass matrix. The equation corresponding to this pattern can be 
expressed in terms of the elements of the neutrino mass matrix as 
\begin{equation}
\label{eq:28}
 (M_{\nu})_{ ee}(M_{\nu})_{ \mu\mu}-(M_{\nu})_{ e\mu}(M_{\nu})_{ e\mu}=0\,.
\end{equation}
Using Eq.~\ref{eq:21}, the two mass ratios for TM$_1$ can be expressed as
\begin{eqnarray}
\label{eq:29} 
&&\frac{m_1}{m_2}=\frac{\mathcal{A}_1\sin2\beta+\mathcal{A}_2\cos2\beta}{(\mathcal{A}_3+\mathcal{A}_4)\sin2(\alpha-\beta)+(\mathcal{A}_5-
\mathcal{A}_6)\cos2(\alpha-\beta)}\,,\nonumber \\ 
&&\frac{m_3}{m_2}=\frac{\mathcal{A}_1\sin2\beta+\mathcal{A}_2\cos2\beta}{\mathcal{A}_7\sin2\alpha+\mathcal{A}_8\cos2\alpha}\,,
\end{eqnarray}
Similarly, for TM$_2$ mixing matrix, the mass ratios can be expressed as
\begin{eqnarray} 
\label{eq:30}
&&\frac{m_1}{m_2}=\frac{(\mathcal{\tilde A}_1+\mathcal{\tilde A}_2)\sin2(\beta-\phi)+(\mathcal{\tilde A}_3+\mathcal{\tilde A}_4)
\cos2(\beta-\phi)}
{\mathcal{\tilde A}_5\sin2(\alpha-\beta)-\mathcal{\tilde A}_6\cos2(\alpha-\beta)}\,,  \nonumber \\
&&\frac{m_3}{m_2}=\frac{(\mathcal{\tilde A}_1+\mathcal{\tilde A}_2)\sin2(\beta-\phi)+(\mathcal{\tilde A}_3+\mathcal{\tilde A}_4)
\cos2(\beta-\phi)}
{\mathcal{\tilde A}_7\sin2(\phi-\alpha)-\mathcal{\tilde A}_8\cos2(\phi-\alpha)}\,.
\end{eqnarray}
All the relevant expressions for $\mathcal{A}_i$ and $\mathcal{\tilde A}_i$ are reported in Eq.~\ref{eq:56} and Eq.~\ref{eq:61} of appendix \ref{app}, 
respectively.
We show the variation of neutrino masses $m_1$, $m_2$ and $m_3$ as a function of $\phi$ in Fig~\ref{fig:3a} and Fig.~\ref{fig:4a} for TM$_1$ 
and TM$_2$ mixing matrix, respectively. It shows normal mass ordering for TM$_1$ mixing matrix while for TM$_2$ mixing matrix, it shows both 
normal and inverted mass ordering. The correlation of $M_{ee}$ and $\sum m_i$ for TM$_1$ and TM$_2$ mixing matrix are shown in 
Fig.~\ref{fig:3b} and Fig.~\ref{fig:4b}, respectively. The vertical red line shows the upper bound of the total neutrino mass reported in 
Ref.~\cite{Zhang:2020mox}. The black, green and blue lines are the experimental upper bounds of the effective Majorana mass as reported in Ref.~\cite{GERDA:2013vls,CUORE:2015hsf,EXO-200:2014ofj,KamLAND-Zen:2012mmx}. In 
Fig.~\ref{fig:3c} and Fig.~\ref{fig:4c}, we have shown the correlation of $M_{\nu}$ with $\sum m_i$ for TM$_1$ and TM$_2$ mixing matrix, 
respectively. It is observed that the total neutrino mass $\sum m_i$ put severe constraint on effective Majorana mass and $M_{\nu}$.  

The range of the absolute neutrino mass scale, the effective Majorana neutrino mass and the effective electron anti-neutrino mass obtained 
for both the mixing matrix are listed in Table.~\ref{tab:3}. The calculated upper bound of $m_{ee}$ is obtained to be of $\mathcal O(10^{-2})$
which is within the reach of neutrinoless double beta decay experiment. The calculated upper bound of $m_{\nu} < 0.06\,{\rm eV}$ may not be 
within the reach of KATRIN experiment. It may, however, be within the reach of next generation experiment such as Project 8. It should, 
however, be mentioned that once the total neutrino mass constraint is imposed, the calculated upper bound of $m_{ee}$ and $m_{\nu}$ is found
to be less than $0.04\,{\rm eV}$.
\begin{table}[htbp]
\begin{tabular}{|c|c|c|c|c|}
 \hline
 Mixing&Mass& $\sum m_i\,{(\rm eV)}$ &$m_{ee}\,{(\rm eV)}$&$m_{\nu}\,{(\rm eV)}$\\
 matrix&ordering& &&\\
\hline
TM$_1$&\textbf{NO} & $[0.059, 0.288]$ & $[1.955\times 10^{-5},  0.048]$&$[8.889 \times 10^{-3}, 0.053]$\\
\hline
\multirow{2}{*}{TM$_2$}&\textbf{NO} & $[0.056, 0.310]$ & $[1.462\times 10^{-5},  0.057]$&$[8.575 \times 10^{-3}, 0.099]$\\ \cline{2-5}
&\textbf{IO} & $[0.094, 0.425]$ & $[0.014, 0.116]$&$[0.044, 0.144]$\\
\hline
\hline
\end{tabular}
\caption{Allowed range of $\sum m_{i}$, $m_{ee}$ and $m_{\nu}$ for C$_{33}$ pattern.}
\label{tab:3}
\end{table}

\subsection{\textbf{Pattern II: $\boldsymbol {C_{22}=0}$}}
\begin{figure}[htbp]
\begin{subfigure}{0.32\textwidth}
\includegraphics[width=\textwidth]{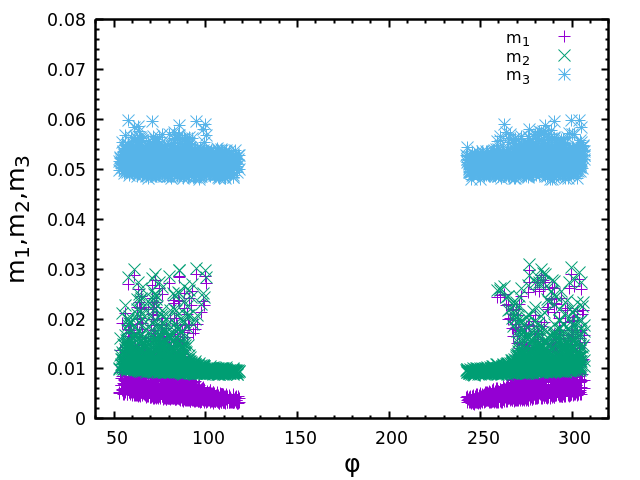}
\caption{}
\label{fig:5a}
\end{subfigure}
\begin{subfigure}{0.32\textwidth}
\includegraphics[width=\textwidth]{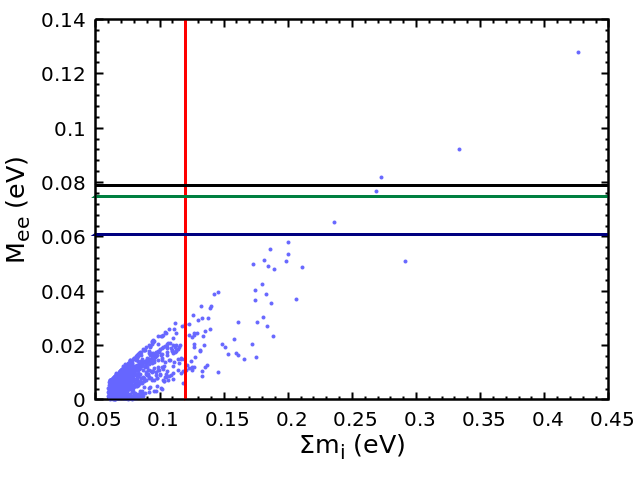}
\caption{}
\label{fig:5b}
\end{subfigure}
\begin{subfigure}{0.32\textwidth}
\includegraphics[width=\textwidth]{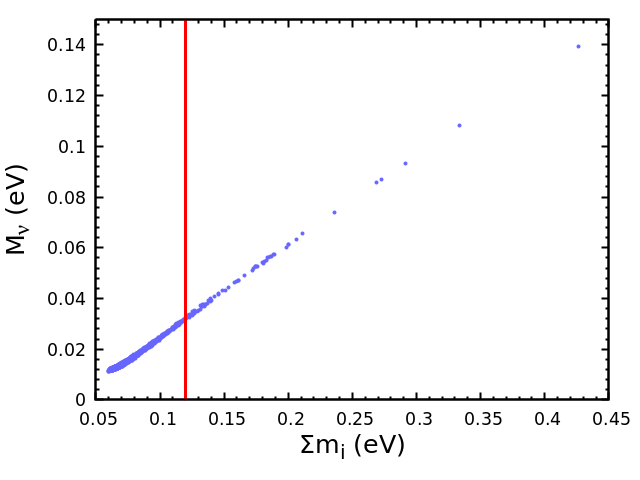}
\caption{}
\label{fig:5c}
\end{subfigure}
\captionsetup{justification=raggedright,singlelinecheck=false}
\caption{Various correlation plots for C$_{22}$ pattern using TM$_1$ mixing matrix. The vertical red line shows the upper bound of the total neutrino mass reported in Ref.~\cite{Zhang:2020mox}. The black, green and blue lines are the experimental upper bounds of the effective Majorana mass reported in Ref.~\cite{GERDA:2013vls,CUORE:2015hsf,EXO-200:2014ofj,KamLAND-Zen:2012mmx}.}
\label{fig:5}
\end{figure}

\begin{figure}[htbp]
\begin{subfigure}{0.32\textwidth}
\includegraphics[width=\textwidth]{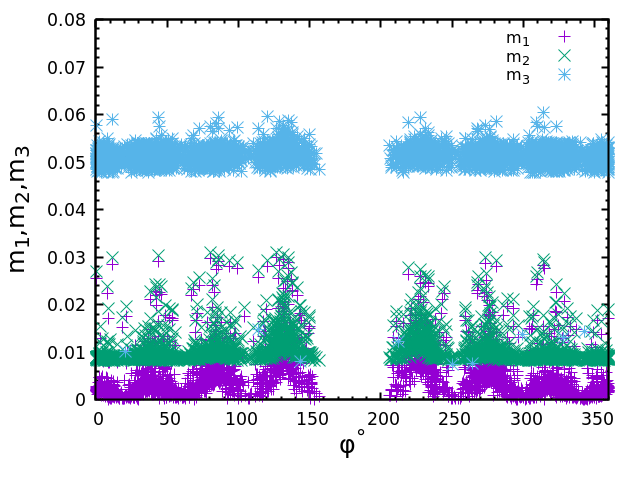}
\caption{}
\label{fig:6a}
\end{subfigure}
\begin{subfigure}{0.32\textwidth}
\includegraphics[width=\textwidth]{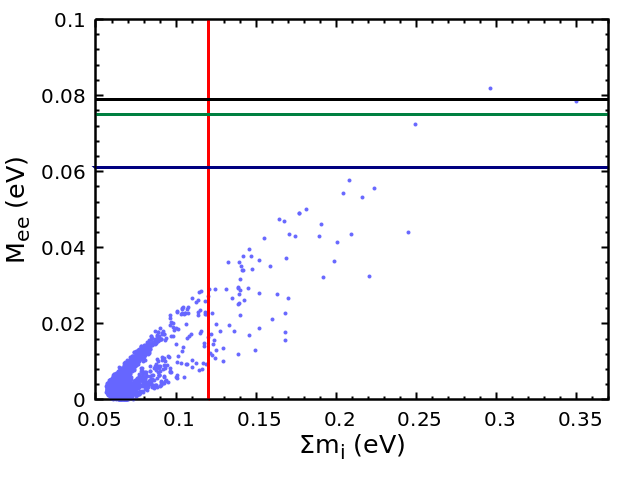}
\caption{}
\label{fig:6b}
\end{subfigure}
\begin{subfigure}{0.32\textwidth}
\includegraphics[width=\textwidth]{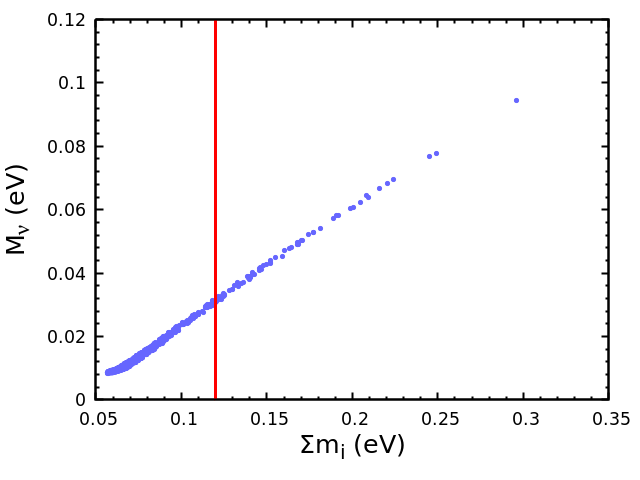}
\caption{}
\label{fig:6c}
\end{subfigure}
\captionsetup{justification=raggedright,singlelinecheck=false}
\caption{Various correlation plots for C$_{22}$ pattern using TM$_2$ mixing matrix. The vertical red line shows the upper bound of the total neutrino mass reported in Ref.~\cite{Zhang:2020mox}. The black, green and blue lines show the experimental upper bounds of the effective Majorana mass reported in Ref.~\cite{GERDA:2013vls,CUORE:2015hsf,EXO-200:2014ofj,KamLAND-Zen:2012mmx}.}
\label{fig:6}
\end{figure}

The vanishing minor condition for this pattern corresponding to element (2,2) is given by
\begin{equation} 
\label{eq:31}
 (M_{\nu})_{ ee}(M_{\nu})_{ \tau\tau}-(M_{\nu})_{ e\tau}(M_{\nu})_{ e\tau}=0\,.
\end{equation}
The two mass ratios for TM$_1$ and TM$_2$ mixing matrix can be expressed as
\begin{eqnarray}
\label{eq:32}
&& \frac{m_1}{m_2}=\frac{\mathcal{B}_1\sin2\beta+\mathcal{B}_2\cos2\beta}{(\mathcal{B}_3+\mathcal{B}_4)\sin2(\alpha-\beta)-
(\mathcal{B}_5-\mathcal{B}_6)\cos2(\alpha-\beta)}\,,\nonumber \\
&&\frac{m_3}{m_2}=\frac{\mathcal{B}_1\sin2\beta+\mathcal{B}_2\cos2\beta}{\mathcal{B}_7\sin2\alpha+\mathcal{B}_8\cos2\alpha}\,,
\end{eqnarray}
and
\begin{eqnarray}
\label{eq:33}
&& \frac{m_1}{m_2}=\frac{(\mathcal{\tilde B}_1+\mathcal{\tilde B}_2)\sin2(\beta-\phi)+(\mathcal{\tilde B}_3+
\mathcal{\tilde B}_4)\cos2(\beta-\phi)}
{\mathcal{\tilde B}_5\sin2(\alpha-\beta)-\mathcal{\tilde B}_6\cos2(\alpha-\beta)}\,,\nonumber \\
&& \frac{m_3}{m_2}=\frac{(\mathcal{\tilde B}_1+\mathcal{\tilde B}_2)\sin2(\beta-\phi)+(\mathcal{\tilde B}_3+\mathcal{\tilde B}_4)\cos2(\beta-\phi)}
{\mathcal{\tilde B}_7\sin2(\phi-\alpha)-\mathcal{\tilde B}_8\cos2(\phi-\alpha)}\,.
\end{eqnarray}
All the relevant expressions for $\mathcal{B}_i$ and $\mathcal{\tilde B}_i$ are reported in Eq.~\ref{eq:57} and Eq.~\ref{eq:62} of appendix \ref{app}.
We show the variation of neutrino masses $m_1$, $m_2$ and $m_3$ as a function of $\phi$ in Fig~\ref{fig:5a} and Fig.~\ref{fig:6a} for TM$_1$
and TM$_2$ mixing matrix, respectively. It shows normal mass ordering for TM$_1$ mixing matrix while for TM$_2$ mixing matrix, it shows both 
normal and inverted mass ordering. The correlation of $M_{ee}$ and $\sum m_i$ for TM$_1$ and TM$_2$ mixing matrix are shown in 
Fig.~\ref{fig:5b} and Fig.~\ref{fig:6b}, respectively. In Fig.~\ref{fig:5c} and Fig.~\ref{fig:6c}, we have shown the correlation of $M_{\nu}$ 
with $\sum m_i$ for TM$_1$ and TM$_2$ mixing matrix, respectively. The phenomenology of this pattern is quite similar to C$_{33}$.

The allowed range of the absolute neutrino mass scale, the effective Majorana mass and the effective electron anti-neutrino mass for this 
pattern are listed in Table.~\ref{tab:4}. 
\begin{table}[htbp!]
\begin{tabular}{|c|c|c|c|c|}
 \hline
 Mixing&Mass& $\sum m_i\,{(\rm eV)}$ &$m_{ee}\,{(\rm eV)}$&$m_{\nu}\,{(\rm eV)}$\\
 matrix&ordering& &&\\
\hline
TM$_1$&\textbf{NO} & $[0.059, 0.425]$ & $[2.097\times 10^{-5},  0.127]$&$[0.011, 0.139]$\\
\hline
\multirow{2}{*}{TM$_2$}&\textbf{NO} & $[0.056, 0.350]$ & $[4.185\times 10^{-6},  0.081]$&$[8.415\times 10^{-3}, 0.113]$\\ \cline{2-5}
&\textbf{IO} & $[0.098, 0.427]$ & $[0.015, 0.082]$&$[0.045, 0.145]$\\
\hline
\hline
\end{tabular}
\caption{Allowed range of $\sum m_{i}$, $m_{ee}$ and $m_{\nu}$ for C$_{22}$ pattern.}
\label{tab:4}
\end{table}
It is evident from Fig.~\ref{fig:5} and Fig.~\ref{fig:6} that the upper bound of $\sum m_i < 1.2\,{\rm eV}$ put severe constraint on the 
value of the effective Majorana mass term $m_{ee}$ and the value of the effective electron anti-neutrino mass $m_{\nu}$. The estimated
upper bound of $m_{ee}$ and $m_{\nu}$ is found to be less than $0.04\,{\rm eV}$.

\subsection{\textbf{Pattern III: $\boldsymbol {C_{31}=0}$}}
\begin{figure}[htbp]
\begin{subfigure}{0.32\textwidth}
\includegraphics[width=\textwidth]{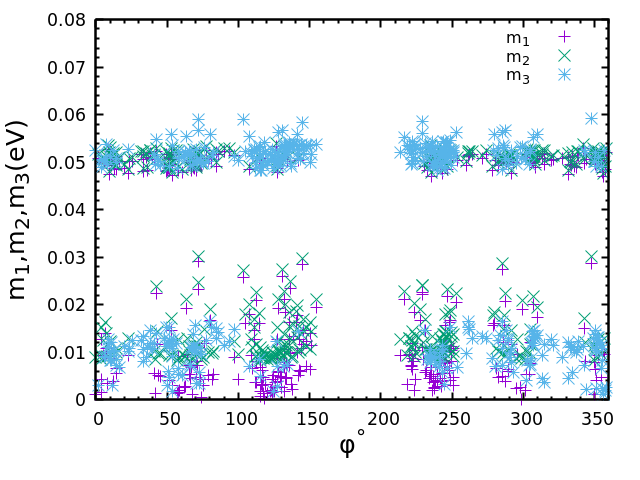}
\caption{}
\label{fig:8a}
\end{subfigure}
\begin{subfigure}{0.32\textwidth}
\includegraphics[width=\textwidth]{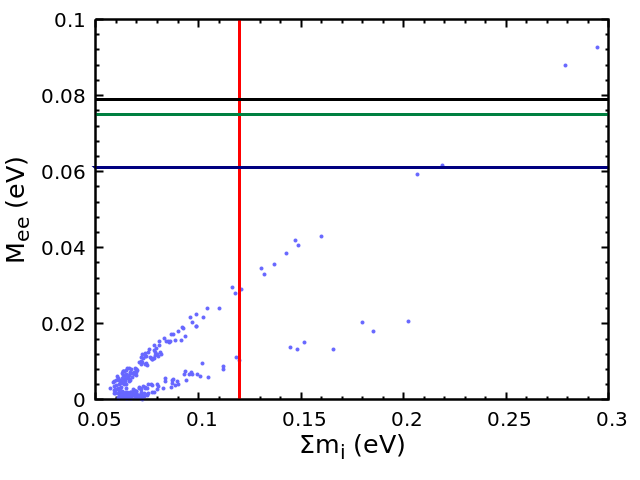}
\caption{}
\label{fig:8b}
\end{subfigure}
\begin{subfigure}{0.32\textwidth}
\includegraphics[width=\textwidth]{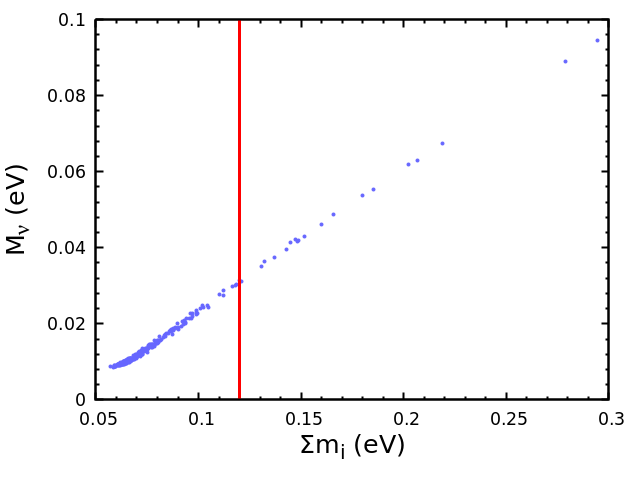}
\caption{}
\label{fig:8c}
\end{subfigure}
\captionsetup{justification=raggedright,singlelinecheck=false}
\caption{Various correlation plots for C$_{31}$ pattern using TM$_2$ mixing matrix. The vertical red line represents the upper bound of the total neutrino mass reported in Ref.~\cite{Zhang:2020mox}. The black, green and blue lines represent the experimental upper bounds of the effective Majorana mass reported in Ref.~\cite{GERDA:2013vls,CUORE:2015hsf,EXO-200:2014ofj,KamLAND-Zen:2012mmx}.}
\label{fig:8}
\end{figure}

This pattern corresponds to the matrix element (3,1) of the neutrino mass matrix. The vanishing minor condition is given by
\begin{equation} 
\label{eq:34} 
 (M_{\nu})_{ e\mu}(M_{\nu})_{ \mu\tau}-(M_{\nu})_{ e\tau}(M_{\nu})_{ \mu\mu}=0\,.
\end{equation}
Using the elements from neutrino mass matrix, one can write 
the neutrino mass ratios for this pattern. With TM$_1$ mixing matrix, we have
\begin{eqnarray}
\label{eq:35}
&& \frac{m_1}{m_2}=\frac{\mathcal{C}_1\sin2\beta+\mathcal{C}_2\cos2\beta}{\mathcal{C}_3\sin2(\alpha-\beta)-\mathcal{C}_4\cos2(\alpha-\beta)},
\nonumber \\
&&\frac{m_3}{m_2}=\frac{\mathcal{C}_1\sin2\beta+\mathcal{C}_2\cos2\beta}{\mathcal{C}_5\sin2\alpha+\mathcal{C}_6\cos2\alpha}\,,
\end{eqnarray}
and for TM$_2$ mixing matrix, we have
\begin{eqnarray}
\label{eq:36}
&& \frac{m_1}{m_2}=\frac{\mathcal{\tilde C}_1 \sin2(\beta-\phi)+\mathcal{\tilde C}_2\cos2(\beta-\phi)}{\mathcal{\tilde C}_3\sin2(\alpha-\beta)-\mathcal{\tilde C}_4\cos2(\alpha-\beta)}\,,\nonumber \\
&& \frac{m_3}{m_2}=\frac{\mathcal{\tilde C}_1\sin2(\beta-\phi)+\mathcal{\tilde C}_2 \cos2(\beta-\phi)}{\mathcal{\tilde C}_5\sin2(\phi-\alpha)+\mathcal{\tilde C}_6\cos2(\phi-\alpha)}\,,
\end{eqnarray}
where all the relevant $\mathcal{C}_i$ and $\mathcal{\tilde C}_i$ are reported in Eq.~\ref{eq:58} and Eq.~\ref{eq:63} of appendix \ref{app}.
We show in Fig.~\ref{fig:7a} and Fig.~\ref{fig:8a} the correlation of neutrino masses $m_1$, $m_2$ and $m_3$ with the unknown parameter $\phi$ for TM$_1$ and TM$_2$ mixing matrix. It shows both 
normal and inverted mass ordering for TM$_1$ and TM$_2$ mixing matrix. Similarly, the correlation of $M_{ee}$ against $\sum m_i$ for TM$_1$ 
and TM$_2$ mixing patterns are shown in Fig.~\ref{fig:7b} and Fig.~\ref{fig:8b},respectively. Moreover, in the Fig.~\ref{fig:7c} and Fig.~\ref{fig:8c}, we have shown the correlation of $M_{\nu}$ with 
$\sum m_i$ for TM$_1$ and TM$_2$, respectively. 
\begin{figure}[htbp]
\begin{subfigure}{0.32\textwidth}
\includegraphics[width=\textwidth]{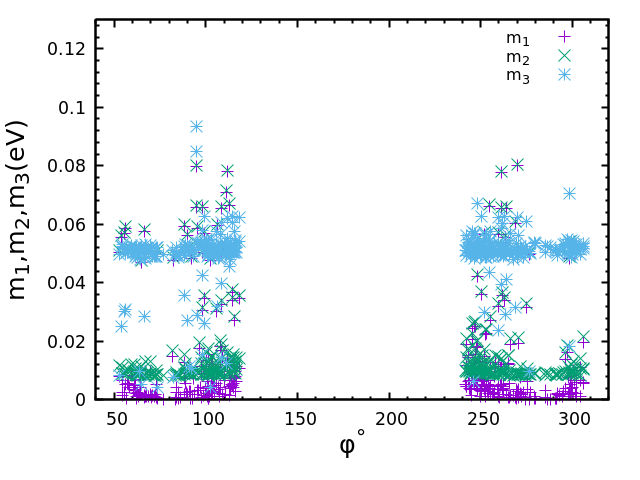}
\caption{}
\label{fig:7a}
\end{subfigure}
\begin{subfigure}{0.32\textwidth}
\includegraphics[width=\textwidth]{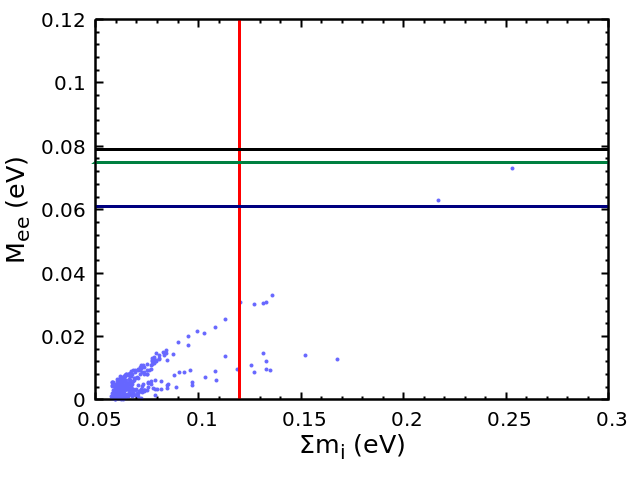}
\caption{}
\label{fig:7b}
\end{subfigure}
\begin{subfigure}{0.32\textwidth}
\includegraphics[width=\textwidth]{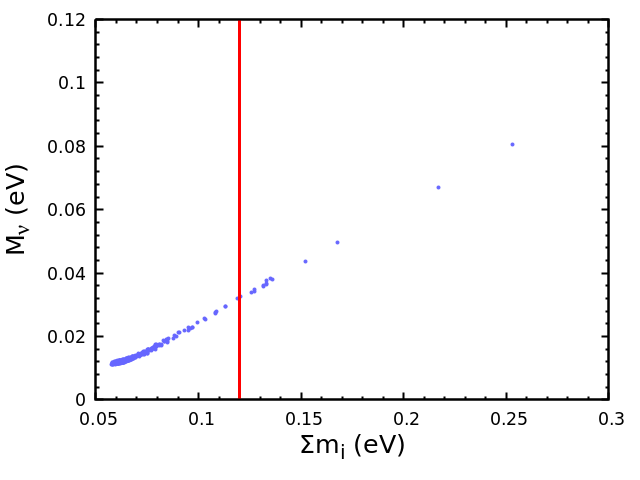}
\caption{}
\label{fig:7c}
\end{subfigure}
\captionsetup{justification=raggedright,singlelinecheck=false}
 \caption{Various correlation plots for C$_{31}$ pattern using TM$_1$ mixing matrix. The vertical red line is the upper bound of the total neutrino mass reported in Ref.~\cite{Zhang:2020mox}. The black, green and blue lines are the experimental upper bounds of the effective Majorana mass reported in Ref.~\cite{GERDA:2013vls,CUORE:2015hsf,EXO-200:2014ofj,KamLAND-Zen:2012mmx}.}
\label{fig:7}
\end{figure}
    
The allowed ranges of the absolute neutrino mass, the effective Majorana mass and the effective electron anti-neutrino mass for both the maxing matrix are listed in the
Table.~\ref{tab:5}.
\begin{table}[htbp]
\begin{tabular}{|c|c|c|c|c|}
 \hline
 Mixing&Mass& $\sum m_i\,{(\rm eV)}$ &$m_{ee}\,{(\rm eV)}$&$m_{\nu}\,{(\rm eV)}$\\
 matrix&ordering& &&\\
\hline
\multirow{2}{*}{TM$_1$}&\textbf{NO} & $[0.057, 0.253]$ & $[1.003\times 10^{-4}, 0.073]$&$[0.011, 0.080$\\ \cline{2-5}
&\textbf{IO} & $[0.096, 0.329]$ & $[0.014, 0.087]$&$[0.044,0.112]$\\
\hline
\multirow{2}{*}{TM$_2$}&\textbf{NO} & $[0.057, 0.294]$ & $[3.658\times 10^{-5}, 0.092]$&$[8.624 \times 10^{-3}, 0.094]$\\ \cline{2-5}
&\textbf{IO} & $[0.092, 0.548]$ & $[0.014, 0.126]$&$[0.045,0.185]$\\
\hline
\hline
\end{tabular}
\caption{Allowed range of $\sum m_{i}$, $m_{ee}$ and $m_{\nu}$ for C$_{31}$ pattern.}
\label{tab:5}
\end{table}
    
\subsection{\textbf{Pattern IV: $\boldsymbol {C_{21}=0}$}}
The vanishing minor condition for this pattern corresponding to element (2,1) of the neutrino mass matrix is given by
\begin{equation}
\label{eq:37}
 (M_{\nu})_{ \mu e}(M_{\nu})_{ \tau\tau}-(M_{\nu})_{\tau e}(M_{\nu})_{ \mu\tau}=0\,.
\end{equation}
The two neutrino mass ratios for this pattern for TM$_1$ and TM$_2$ mixing matrix are given by
\begin{eqnarray}
\label{eq:38}
&& \frac{m_1}{m_2}=\frac{\mathcal{D}_1\sin2\beta+\mathcal{D}_2\cos2\beta}{\mathcal{D}_3\sin2(\alpha-\beta)
 +\mathcal{D}_4\cos2(\alpha-\beta)},\nonumber \\
&&\frac{m_3}{m_2}=\frac{\mathcal{D}_1\sin2\beta+\mathcal{D}_2\cos2\beta}{\mathcal{D}_5\sin2\alpha+\mathcal{D}_6\cos2\alpha}\,,
\end{eqnarray}
and
\begin{eqnarray}
\label{eq:39}
&& \frac{m_1}{m_2}=\frac{\mathcal{\tilde D}_1 \sin2(\beta-\phi)+\mathcal{\tilde D}_2\cos2(\beta-\phi)}{\mathcal{\tilde D}_3\sin2(\alpha-\beta)+\mathcal{\tilde D}_4\cos2(\alpha-\beta)}\,,\nonumber \\
&& \frac{m_3}{m_2}=\frac{\mathcal{\tilde D}_1\sin2(\beta-\phi)+\mathcal{\tilde D}_2 \cos2(\beta-\phi)}{\mathcal{\tilde D}_5\sin2(\phi-\alpha)-\mathcal{\tilde D}_6\cos2(\phi-\alpha)}\,,
\end{eqnarray}
where all the relevant $\mathcal{D}_i$ and $\mathcal{\tilde D}_i$ are reported in Eq.~\ref{eq:59} and Eq.~\ref{eq:64} of appendix \ref{app}. In Fig.~\ref{fig:9a} and Fig.~\ref{fig:10a} we have shown the correlation of neutrino masses $m_1$, $m_2$ and $m_3$ with the unknown parameter $\phi$ for TM$_1$ and TM$_2$ mixing matrix. It shows both 
normal and inverted mass ordering for TM$_1$ and TM$_2$ mixing matrix. The correlation of $M_{ee}$ against $\sum m_i$ for TM$_1$ and TM$_2$ mixing patterns are shown in Fig.~\ref{fig:9b} and Fig.~\ref{fig:10b},respectively. In the Fig.~\ref{fig:9c} and Fig.~\ref{fig:10c}, we have shown the correlation of $M_{\nu}$ with $\sum m_i$ for TM$_1$ and TM$_2$, respectively.

\begin{figure}[htbp]
\begin{subfigure}{0.32\textwidth}
\includegraphics[width=\textwidth]{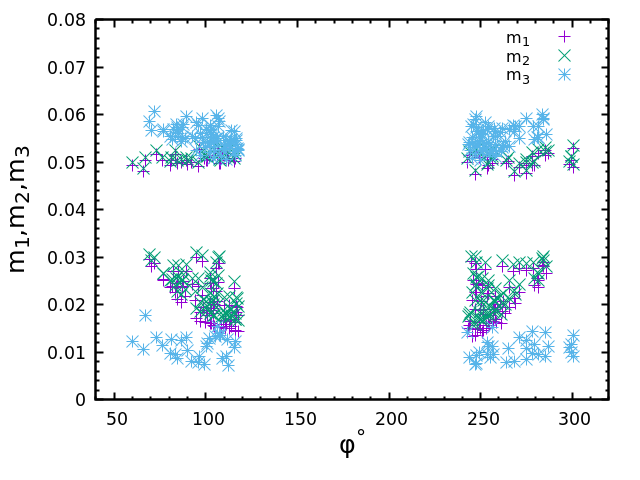}
\caption{}
\label{fig:9a}
\end{subfigure}
\begin{subfigure}{0.32\textwidth}
\includegraphics[width=\textwidth]{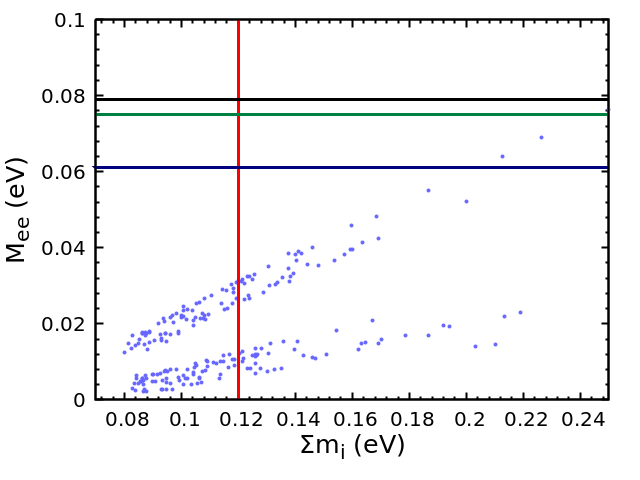}
\caption{}
\label{fig:9b}
\end{subfigure}
\begin{subfigure}{0.32\textwidth}
\includegraphics[width=\textwidth]{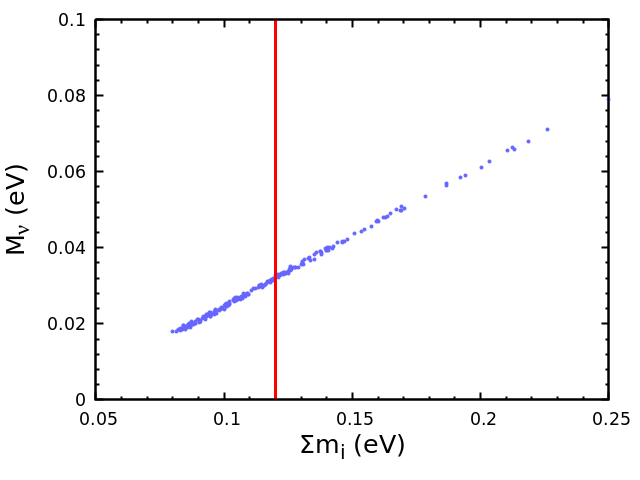}
\caption{}
\label{fig:9c}
\end{subfigure}
\captionsetup{justification=raggedright,singlelinecheck=false}
\caption{Various correlation plots for C$_{21}$ pattern using TM$_1$ mixing matrix. The vertical red line is the upper bound of the total neutrino mass reported in Ref.~\cite{Zhang:2020mox}. The black, green and blue lines are the experimental upper bounds of the effective Majorana mass reported in Ref.~\cite{GERDA:2013vls,CUORE:2015hsf,EXO-200:2014ofj,KamLAND-Zen:2012mmx}.}
\label{fig:9}
\end{figure}
    
\begin{figure}
\begin{subfigure}{0.32\textwidth}
\includegraphics[width=\textwidth]{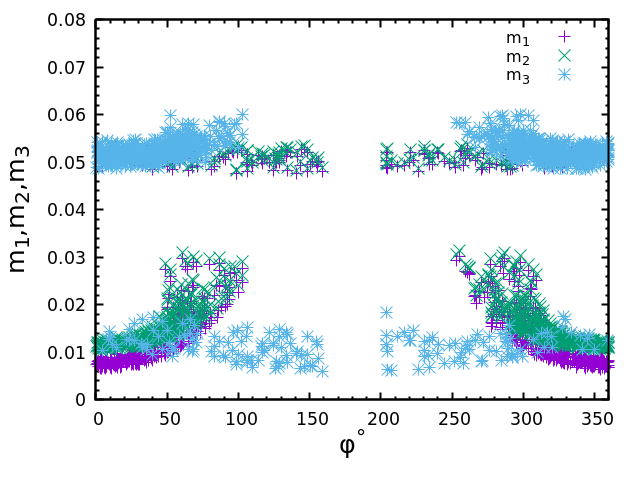}
\caption{}
\label{fig:10a}
\end{subfigure}
\hfill
\begin{subfigure}{0.32\textwidth}
\includegraphics[width=\textwidth]{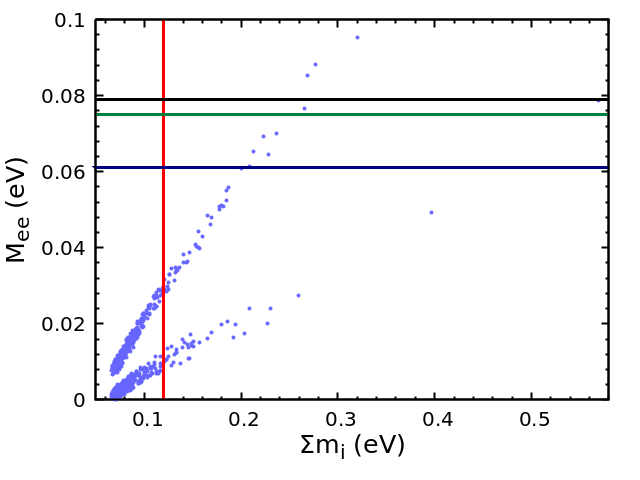}
\caption{}
\label{fig:10b}
\end{subfigure}
\hfill
\begin{subfigure}{0.32\textwidth}
\includegraphics[width=\textwidth]{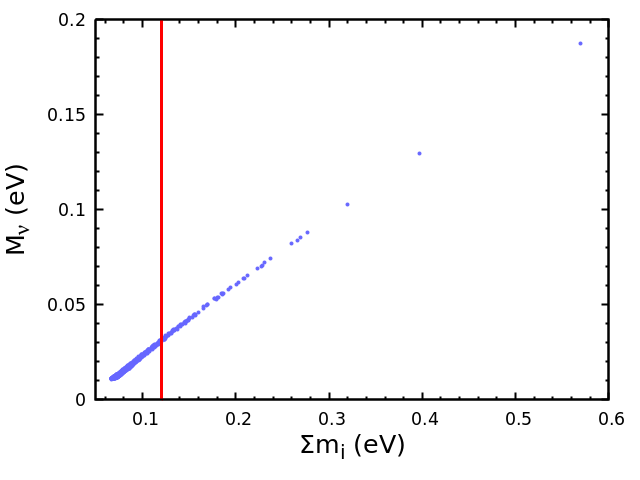}
\caption{}
\label{fig:10c}
\end{subfigure}
\captionsetup{justification=raggedright,singlelinecheck=false}
\caption{Various correlation plots for C$_{21}$ pattern using TM$_2$ mixing matrix. The vertical red line shows the upper bound of the total neutrino mass reported in Ref.~\cite{Zhang:2020mox}. The black, green and blue lines show the experimental upper bounds of the effective Majorana mass reported in Ref.~\cite{GERDA:2013vls,CUORE:2015hsf,EXO-200:2014ofj,KamLAND-Zen:2012mmx}.}
\label{fig:10}
\end{figure}

We also report the allowed ranges of the absolute neutrino mass, the effective Majorana mass and the effective electron anti-neutrino mass for both the maxing matrix
in Table.~\ref{tab:6}. The phenomenology of this pattern is very similar to that of $C_{31}$.

\begin{table}[htbp]
\begin{tabular}{|c|c|c|c|c|}
 \hline
 Mixing&Mass& $\sum m_i\,{(\rm eV)}$ &$m_{ee}\,{(\rm eV)}$&$m_{\nu}\,{(\rm eV)}$\\
 matrix&ordering&& &\\
\hline
\multirow{2}{*}{TM$_1$}&\textbf{NO} & $[0.080, 0.249]$ & $[2.213\times 10^{-3}, 0.076]$&$[0.017, 0.079]$\\ \cline{2-5}
&\textbf{IO} & $[0.101, 0.720]$ & $[0.014, 0.125]$&$[0.044, 0.241]$\\
\hline
\multirow{2}{*}{TM$_2$}&\textbf{NO} & $[0.066, 0.569]$ & $[1.655\times 10^{-4}, 0.095]$&$[0.010, 0.187]$\\ \cline{2-5}
&\textbf{IO} & $[0.097, 0.458]$ & $[0.015, 0.137]$&$[0.045, 0.154]$\\
\hline
\hline
\end{tabular}
\captionof{table}{Allowed range of $\sum m_{i}$, $m_{ee}$ and $m_{\nu}$ for C$_{21}$ pattern.}
\label{tab:6}
\end{table}

\subsection{\textbf{Pattern V: $\boldsymbol {C_{32}=0}$}}
\begin{figure}[htbp]
\begin{subfigure}{0.32\textwidth}
\includegraphics[width=\textwidth]{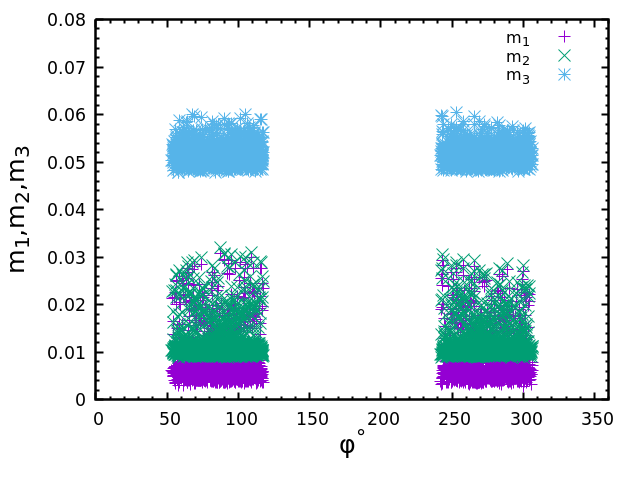}
\caption{}
\label{fig:11a}
\end{subfigure}
\begin{subfigure}{0.32\textwidth}
\includegraphics[width=\textwidth]{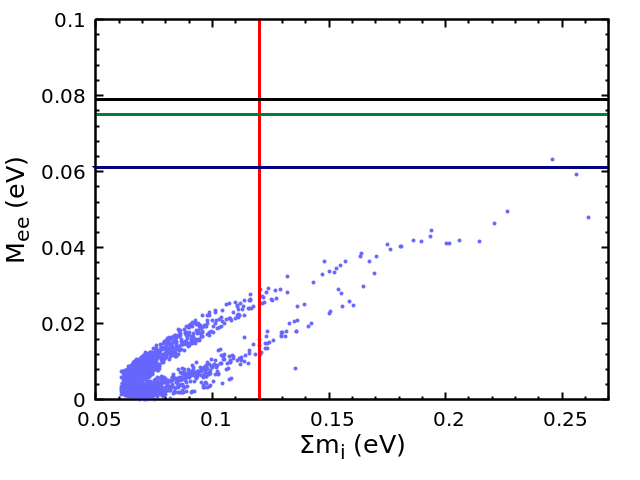}
\caption{}
\label{fig:11b}
\end{subfigure}
\begin{subfigure}{0.32\textwidth}
\includegraphics[width=\textwidth]{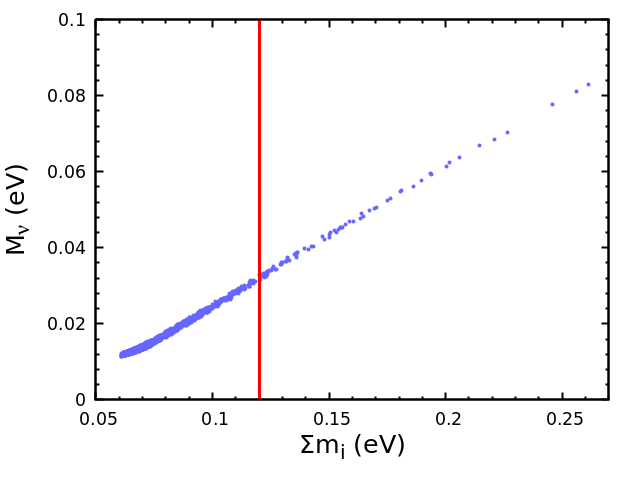}
\caption{}
\label{fig:11c}
\end{subfigure}
\captionsetup{justification=raggedright,singlelinecheck=false}
\caption{Various correlation plots for C$_{32}$ pattern using TM$_1$ mixing matrix. The vertical red line represents the upper bound of the total neutrino mass reported in Ref.~\cite{Zhang:2020mox}. The black, green and blue lines represent the experimental upper bounds of the effective Majorana mass reported in Ref.~\cite{GERDA:2013vls,CUORE:2015hsf,EXO-200:2014ofj,KamLAND-Zen:2012mmx}.}
\label{fig:11}
\end{figure}

\begin{figure}[htbp]
\begin{subfigure}{0.32\textwidth}\includegraphics[width=\textwidth]{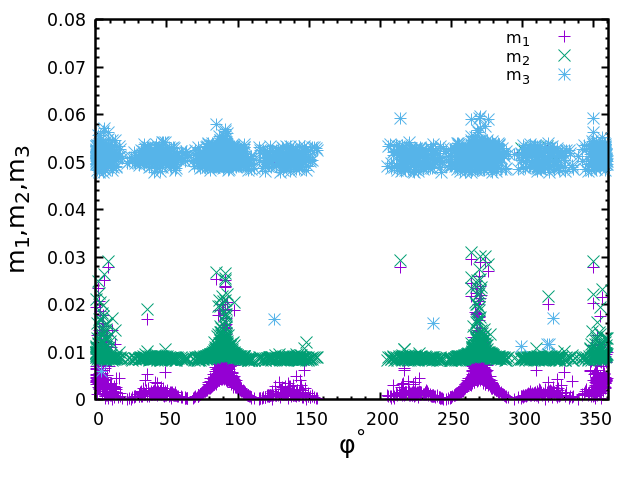}\caption{}
\label{fig:12a}
\end{subfigure}
\begin{subfigure}{0.32\textwidth}
\includegraphics[width=\textwidth]{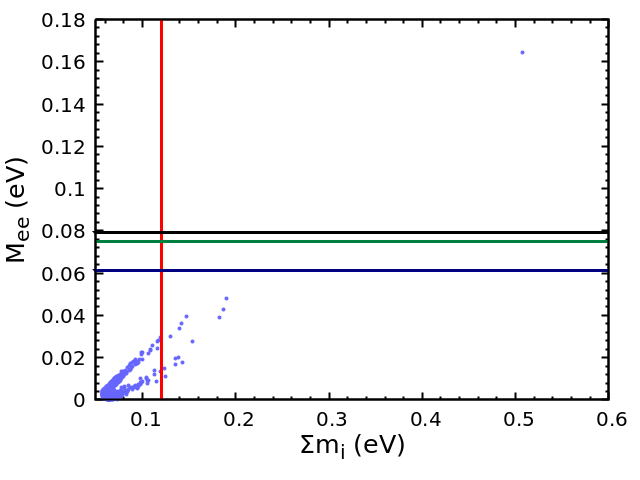}
\caption{}
\label{fig:12b}
\end{subfigure}
\begin{subfigure}{0.32\textwidth}
\includegraphics[width=\textwidth]{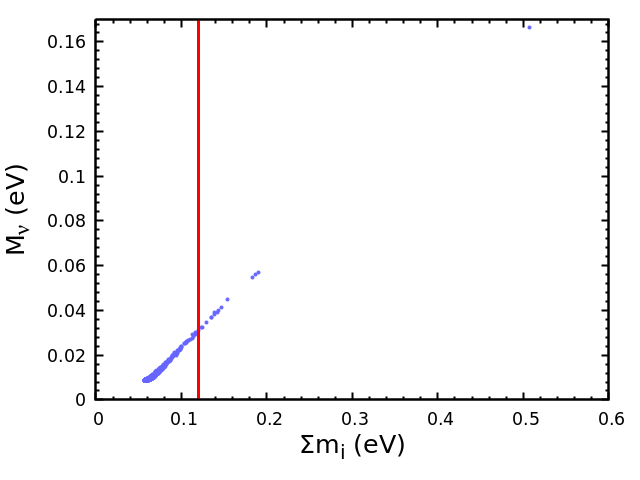}
\caption{}
\label{fig:12c}
\end{subfigure}
\captionsetup{justification=raggedright,singlelinecheck=false}
\caption{Various correlation plots for C$_{32}$ pattern using TM$_2$ mixing matrix. The vertical red line is the upper bound of the total neutrino mass reported in Ref.~\cite{Zhang:2020mox}. The black, green and blue lines are the experimental upper bounds of the effective Majorana mass reported in Ref.~\cite{GERDA:2013vls,CUORE:2015hsf,EXO-200:2014ofj,KamLAND-Zen:2012mmx}.}
\label{fig:12}
\end{figure}

The condition of vanishing minor for this pattern is given by
\begin{equation} 
\label{eq:40}
 (M_{\nu})_{ ee}(M_{\nu})_{ \mu\tau}-(M_{\nu})_{\mu e}(M_{\nu})_{ e\tau}=0.
\end{equation}
For this pattern, the two neutrino mass ratios for TM$_1$ and TM$_2$ mixing matrix are given by
\begin{eqnarray}
\label{eq:41}
&& \frac{m_1}{m_2}=\frac{\mathcal{E}_1\sin2\beta+\mathcal{E}_2\cos2\beta}{\mathcal{E}_3\sin2(\alpha-\beta)
 +\mathcal{E}_4\cos2(\alpha-\beta)}\,,\nonumber \\
&&\frac{m_3}{m_2}=\frac{\mathcal{E}_1\sin2\beta+\mathcal{E}_2\cos2\beta}{\mathcal{E}_5\sin2\alpha+\mathcal{E}_6\cos2\alpha},
\end{eqnarray}
and 
\begin{eqnarray}
\label{eq:42}
&& \frac{m_1}{m_2}=\frac{\mathcal{\tilde E}_1 \sin2(\beta-\phi)+\mathcal{\tilde E}_2\cos2(\beta-\phi)}{\mathcal{\tilde E}_3\sin2(\alpha-\beta)+\mathcal{\tilde E}_4\cos2(\alpha-\beta)}\,,\nonumber \\
&& \frac{m_3}{m_2}=\frac{\mathcal{\tilde E}_1\sin2(\beta-\phi)+\mathcal{\tilde E}_2 \cos2(\beta-\phi)}{\mathcal{\tilde E}_5\sin2(\phi-\alpha)-\mathcal{\tilde E}_6\cos2(\phi-\alpha)}\,,
\end{eqnarray}
respectively. The relevant expressions for $\mathcal{E}_i$ and $\mathcal{\tilde E}_i$ are are reported in Eq.~\ref{eq:60} and Eq.~\ref{eq:65} of appendix \ref{app}. The correlation of neutrino masses $m_1$, $m_2$ and $m_3$ with the unknown parameter $\phi$ are shown in Fig.~\ref{fig:11a} and Fig.~\ref{fig:12a}, respectively for TM$_1$ and TM$_2$ mixing matrix. It is observed that, it shows normal mass ordering for TM$_1$ mixing matrix, whereas, for TM$_2$ mixing matrix, it shows both normal and inverted mass ordering. The correlation of $M_{ee}$ and $\sum m_i$ are shown in Fig.~\ref{fig:11b} and Fig.~\ref{fig:12b}, respectively using TM$_1$ and TM$_2$ mixing matrix. In Fig.~\ref{fig:11c} and Fig.~\ref{fig:12c}, we have shown the correlation of $M_{\nu}$ with $\sum m_i$  using TM$_1$ and TM$_2$ mixing matrix, respectively. 

The allowed ranges of all the relevant parameters such as the  absolute neutrino mass, the effective Majorana mass and the effective electron anti-neutrino mass under normal and 
inverted ordering for both the mixing matrix are reported in Table.~\ref{tab:7}.
\begin{table}[htbp!]
\begin{tabular}{|c|c|c|c|c|}
 \hline
 Mixing&Mass& $\sum m_i\,{(\rm eV)}$ &$m_{ee}\,{(\rm eV)}$&$m_{\nu}\,{(\rm eV)}$\\
 matrix&ordering&& &\\
\hline
TM$_1$&\textbf{NO} & $[0.061, 0.261]$ & $[1.309\times 10^{-4}, 0.063$&$[0.011, 0.082]$\\
\hline
\multirow{2}{*}{TM$_2$}&\textbf{NO} & $[0.056, 0.507]$ & $[6.054\times 10^{-5}, 0.164]$&$[0.084, 0.166]$\\ \cline{2-5}
&\textbf{IO} & $[0.101, 0.346]$ & $[0.017, 0.078]$&$[0.045, 0.118]$\\
\hline
\hline
\end{tabular}
\caption{Allowed range of $\sum m_{i}$, $m_{ee}$ and $m_{\nu}$ for C$_{32}$ pattern.}
\label{tab:7}
\end{table}

\subsection{\textbf{Pattern VI: $\boldsymbol {C_{11}=0}$}}
\begin{figure}[htbp!]
\begin{subfigure}{0.32\textwidth}
\includegraphics[width=\textwidth]{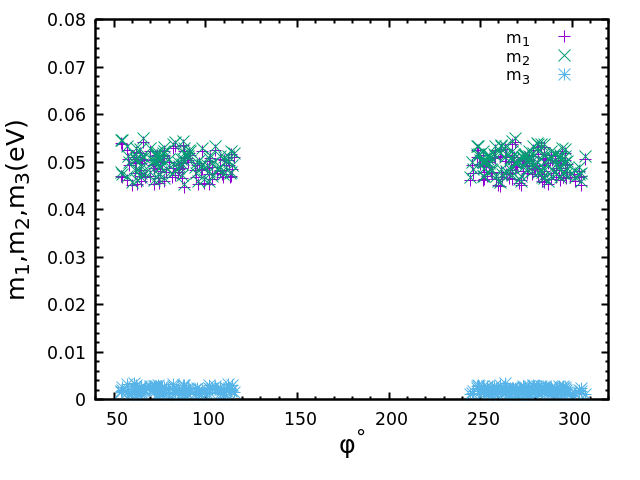}
\caption{}
\label{fig:13a}
\end{subfigure}
\begin{subfigure}{0.32\textwidth}
\includegraphics[width=\textwidth]{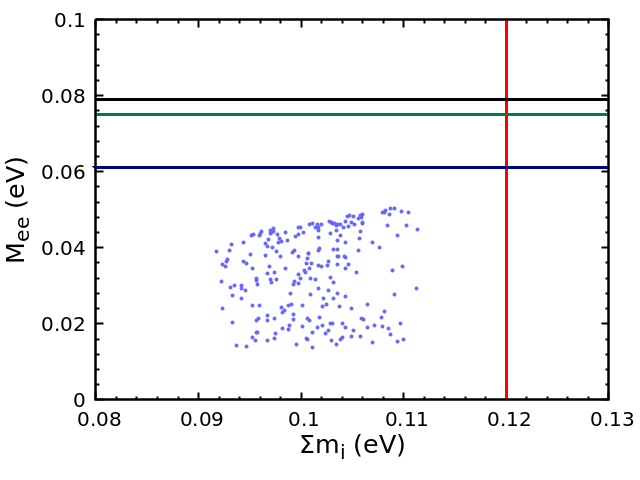}
\caption{}
\label{fig:13b}
\end{subfigure}
\begin{subfigure}{0.32\textwidth}
\includegraphics[width=\textwidth]{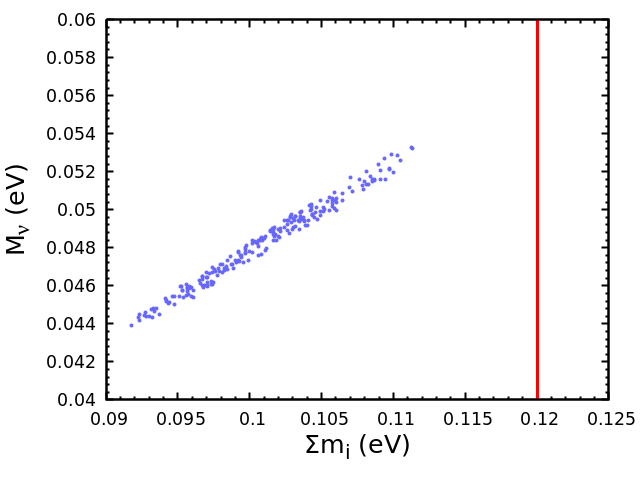}
\caption{}
\label{fig:13c}
\end{subfigure}
\captionsetup{justification=raggedright,singlelinecheck=false}
\caption{Various correlation plots for C$_{11}$ pattern using TM$_1$ mixing matrix. The vertical red line represents the upper bound of the total neutrino mass reported in Ref.~\cite{Zhang:2020mox}. The black, green and blue lines represent the experimental upper bounds of the effective Majorana mass reported in Ref.~\cite{GERDA:2013vls,CUORE:2015hsf,EXO-200:2014ofj,KamLAND-Zen:2012mmx}.}
\label{fig:13}
\end{figure}

\begin{figure}[htbp!]
\begin{subfigure}{0.32\textwidth}
\includegraphics[width=\textwidth]{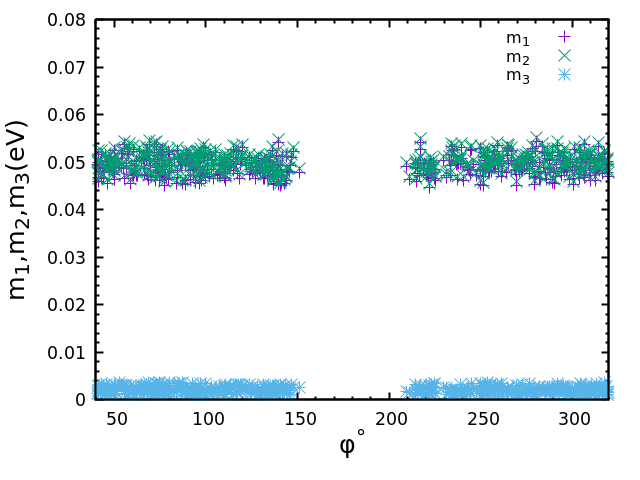}
\caption{}
\label{fig:14a}
\end{subfigure}
\begin{subfigure}{0.32\textwidth}
\includegraphics[width=\textwidth]{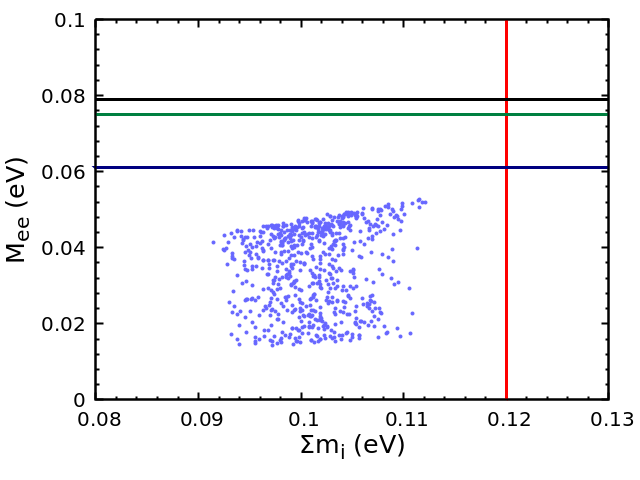}
\caption{}
\label{fig:14b}
\end{subfigure}
\begin{subfigure}{0.32\textwidth}
\includegraphics[width=\textwidth]{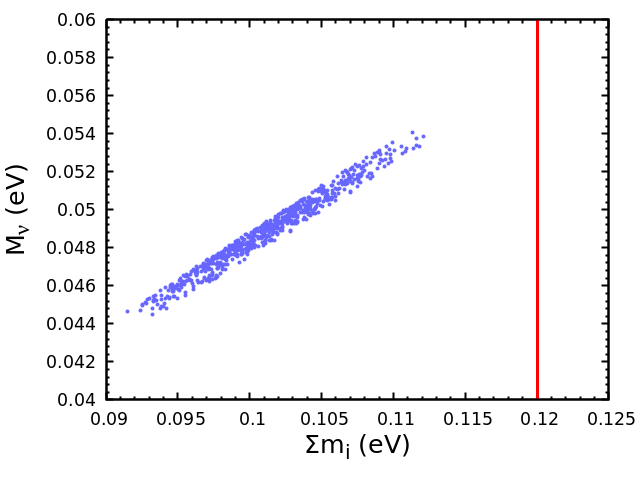}
\caption{}
\label{fig:14c}
\end{subfigure}
\captionsetup{justification=raggedright,singlelinecheck=false}
\caption{Various correlation plots for C$_{11}$ pattern using TM$_2$ mixing matrix. The vertical red line is the upper bound of the total neutrino mass reported in Ref.~\cite{Zhang:2020mox}. The black, green and blue lines are the experimental upper bounds of the effective Majorana mass reported in Ref.~\cite{GERDA:2013vls,CUORE:2015hsf,EXO-200:2014ofj,KamLAND-Zen:2012mmx}.}
\label{fig:14}
\end{figure}

The vanishing minor condition for this pattern is given by
\begin{equation} 
\label{eq:43}
 (M_{\nu})_{\mu \mu}(M_{\nu})_{ \tau\tau}-(M_{\nu})_{\tau \mu}(M_{\nu})_{ \mu\tau}=0\,.
\end{equation}
The two neutrino mass ratios can be obtained using the elements from neutrino mass matrix. For TM$_1$ mixing matrix, we have 
\begin{eqnarray} 
\label{eq:44}
&& \frac{m_1}{m_2}=\frac{2 \sin2\beta}{\cos^2\theta \sin2(\alpha-\beta)}\,,\nonumber \\ 
&&\frac{m_3}{m_2}=-\frac{\tan^2\theta \sin2\beta}{\sin2\alpha}\,,
\end{eqnarray}
and for TM$_2$ mixing matrix, we have
\begin{eqnarray} 
\label{eq:46}
&& \frac{m_1}{m_2}=\frac{2\cos^2\theta \sin2\beta}{\sin2(\alpha-\beta)}\,,\nonumber \\ 
&&\frac{m_3}{m_2}=-\frac{2\sin^2\theta \sin2\beta}{\sin2\alpha}\,.
\end{eqnarray}
Using Eq.~\ref{eq:44}, we obtain the mass relation for TM$_1$ mixing matrix as
\begin{eqnarray} 
\label{eq:49}
 m_{1}\sin2(\alpha-\beta)-2 m_{2}\sin2\beta+m_{3}\sin2\alpha =0\,
\end{eqnarray}
and using Eq.~\ref{eq:46}, we obtain the mass relation for TM$_2$ mixing matrix as
\begin{eqnarray} 
 m_{1}\sin2(\alpha-\beta)-2 m_{2}\sin2\beta-m_{3}\sin2\alpha =0\,.
\end{eqnarray}
This pattern gives a clear inverted mass ordering for both TM$_1$ and TM$_2$ mixing matrix. The correlation of the  neutrino masses 
$m_1$, $m_2$ and $m_3$ for both the mixing patterns with the unknown parameter $\phi$ is shown in Fig.~\ref{fig:13a} and Fig.~\ref{fig:14a}, 
respectively. The correlation of $M_{ee}$ with $\sum m_i$ for TM$_1$ and TM$_2$ are shown in Fig.~\ref{fig:13b} and Fig.~\ref{fig:14b}, 
respectively. In Fig.~\ref{fig:13c} and Fig.~\ref{fig:14c}, we have shown the 
correlation of $M_{\nu}$ with $\sum m_i$ for TM$_1$ and TM$_2$ mixing matrix, respectively. 

The allowed ranges of the absolute neutrino mass, the effective Majorana mass and the effective electron anti-neutrino mass obtained for both the mixing matrix are listed in the tables ~\ref{tab:8}.

\begin{table}[htbp]
\begin{tabular}{|c|c|c|c|c|}
 \hline
 Mixing&Mass& $\sum m_i\,{(\rm eV)}$ &$m_{ee}\,{(\rm eV)}$&$m_{\nu}\,{(\rm eV)}$\\
 matrix&ordering& &&\\
\hline
TM$_1$&\textbf{IO} & $[0.092, 0.111]$ & $[0.013, 0.050]$&$[0.043, 0.053]$\\
\hline
TM$_2$&\textbf{IO} & $[0.091, 0.112]$ & $[0.014, 0.052]$&$[0.044, 0.054$\\ 
\hline
\hline
\end{tabular}
\caption{Allowed range of $\sum m_{i}$, $m_{ee}$ and $m_{\nu}$ for C$_{11}$ pattern.}
\label{tab:8}
\end{table}

\section{Degree of Fine tuning in the neutrino mass matrix}
\label{section:6}
In this section, we wish to determine whether the entries of the neutrino mass matrix are fine tuned or not. 
In order to determine the degree of fine tuning of the mass matrix
elements, we define a parameter $d_{FT}$ ~\cite{Altarelli:2010at,Meloni:2012sx} which is obtained as the sum of the absolute values of the 
ratios between each parameter and its
error. We follow Ref.~\cite{Altarelli:2010at,Meloni:2012sx} and define the fine tuning parameter as
\begin{equation}
 d_{FT}=\sum \Big| \frac{par_i}{err_i}\Big|\,,
\end{equation}
where $par_i = (\theta_{12}, \theta_{13}, \theta_{23}, \Delta m_{21}^2, \Delta m_{31}^2)$ is the best fit values of the parameters. 
The error $err_i$ for each parameter is obtained from the shift of best fit value that changes the $\chi^2_{min}$
value by one unit keeping all other parameters fixed at their best fit values. 
To determine the best fit values of all the parameters, we perform a $\chi^2$ analysis of all the classes of one minor zero and find the 
$\chi^2_{min}$. We define the $\chi^2$ as follows:
\begin{equation}
  \chi^2=\sum\limits_{i=1}^{3} \frac{\Big(\theta_{i}^{cal}- \theta_{i}^{exp}\Big)^2}{(\sigma_{i}^{exp})^2}+\sum\limits_{j=21,3l} \frac{\Big(\Delta m_{j}^{cal}- \Delta m_{j}^{exp}\Big)^2}{(\sigma_{j}^{exp})^2}\,,
 \end{equation}
where $\theta_i=(\theta_{12},\theta_{13},\theta_{23})$ and $\Delta m_{j}= (\Delta m_{21}^2,\,\Delta m_{31}^2)$. Here $\theta_{i}^{cal}$ and 
$\Delta m_{j}^{cal} $ represent the theoretical value of 
$\theta_{i}$ and $\Delta m_{j}$, respectively, whereas  $\theta_{i}^{exp}$ and $\Delta m_{j}^{exp} $
represent measured central value of $\theta_{i}$ and $\Delta m_{j}$, respectively. 
It should be noted that $\theta_{i}^{cal}$ and  $\Delta m_{j}^{cal}$ depend on four unknown model parameters, namely $\theta$, $\phi$, 
$\alpha$ and $\beta$.
Similarly, the uncertainties in the measured value of 
$\theta_{i}$ and $\Delta m_{j}$ are represented by $\sigma_{i}^{exp}$ and $\sigma_{j}^{exp}$, respectively. The central values and the
corresponding uncertainties in each parameter are reported in Table.~\ref{tab:2}.

We first compute $d_{Data}$ which is defined as the sum of the absolute values of the ratios between the measured values
of each parameter and its error from Table.~\ref{tab:2}. We obtain the value of $d_{Data}$ to be around $200$ for both normal and inverted 
ordering case.
The degree of fine tuning can be roughly estimated from the value of $d_{FT}$ because if the $d_{FT}$ value is large then a minimal variation 
of the corresponding parameters give large difference on the value of $\chi^2$. Hence a large value of $d_{FT}$ corresponds to a strong fine
tuning of the mass matrix elements and vice versa. The $\chi_{min}^2$ value and the corresponding best fit values of the unknown parameters
of the neutrino mass matrix $\theta$, $\phi$, $\alpha$, $\beta$ and the value of $d_{FT}$ parameter for each patterns are listed in the 
Table.~\ref{tab:9} and Table.~\ref{tab:10} for the TM$_{1}$ and TM$_2$ mixing matrix respectively. We also report the best fit values of 
several 
observables such as $\theta_{12}$, $\theta_{13}$, $\theta_{23}$, $\Delta m_{21}^2$ and $\Delta m_{31}^2$ for each pattern. For the patterns 
$C_{33},C_{22},C_{31},C_{32}$ and $C_{21}$, the results are for NO case and for the pattern $C_{11}$, the results are for IO case. As the 
pattern $C_{11}$ follows the IO, the $\chi_{min}^2$ value obtained for this pattern is large for both TM$_1$ and TM$_2$ mixing matrix.  
The best fit values of the mixing angles $\theta_{23}$, $\theta_{12}$, $\theta_{13}$ and the mass squared differences $\Delta m^2_{21}$, 
$\Delta m^2_{31}$ obtained for each pattern are compatible with the experimentally measured values reported in Table.~\ref{tab:2}.

\begin{table}[htbp]
\begin{tabular}{|c|c|c|c|c|c|c|c|c|c|c|c|}
 \hline
Type& $\chi^2_{min}$ & $d_{FT}$&$\theta_{13}^\circ $ &$\theta_{12}^\circ $ &$\theta_{23}^\circ $ &$\theta^\circ $ & $\phi^\circ$&$\alpha^\circ $&$\beta^\circ$ & $\Delta m^2_{21}\,{(10^{-5}\rm eV^2)}$ & $\Delta m^2_{3l}\,{(10^{-3}\rm eV^2)}$\\
\hline
\textbf{$C_{33}$}&$2.66$ &$8.38\times$$10^3$&$8.48$ &$34.35$ & $48.59$&$14.80$ &$287.67$&$290.89$&$320.72$&$7.51$ &$2.42$\\

\hline
\textbf{$C_{22}$}&$2.80$ &$2.35\times$$10^5$ &$ 8.46$ & $34.36$ & $48.77 $&$14.77$& $251.38$&$313.24$&$74.37$&$ 7.48$ &$2.45$\\

\hline
\textbf{$C_{31}$}&$1.69$ &$8.43\times$$10^3$&$ 8.57 $& $34.33$& $48.83$ &$14.96$& $108.69$&$7.46$&$338.00$&$7.47$&$2.44$\\

\hline
\textbf{$C_{32}$}&$1.69$ &$83.72$& $8.60$ &$ 34.33$ & $49.56$ & $15.02$& $76.68$&$84.95$&$357.14$&$7.47$&$2.44$\\

\hline
\textbf{$C_{21}$}&$3.24$ &$ 1.36\times$$10^5$& $8.67$&$ 34.31$ & $50.00 $&$15.07$&$63.50$&$269.82$&$216.81$&$7.47$& $2.41$\\

\hline
\textbf{$C_{11}$}& $4.05$&$ 4.88\times$$10^2$& $8.41$ & $34.37$&$ 49.33$ &$14.68$&$ 291.60$&$188.42$&$83.27$& $7.40$&$-2.46$\\
\hline
\end{tabular}
\caption{The values of $\chi_{min}^2$, $d_{FT}$, the best fit values of $\theta_{13}^\circ $, $\theta_{12}^\circ $, $\theta_{23}^\circ $, $\theta^\circ $, $\phi^\circ$,$\alpha^\circ $, $\beta^\circ$, $\Delta m^2_{21}\,{(10^{-5}\rm eV^2)}$ and $\Delta m^2_{3l}\,{(10^{-3}\rm eV^2)}$ for TM$_{1}$ mixing matrix.}
\label{tab:9}
\end{table} 

\begin{table}[htbp]
\begin{tabular}{|c|c|c|c|c|c|c|c|c|c|c|c|}
 \hline
Type& $\chi^2_{min}$ & $d_{FT}$&$\theta_{13}^\circ $ &$\theta_{12}^\circ $ &$\theta_{23}^\circ $ &$\theta^\circ $ & $\phi^\circ$&$\alpha^\circ $&$\beta^\circ$ & $\Delta m^2_{21}\,{(10^{-5}\rm eV^2)}$ & $\Delta m^2_{3l}\,{(10^{-3}\rm eV^2)}$\\
\hline
\textbf{$C_{33}$}&$8.87$ &$5.86\times$$10^2$& $8.57$&$ 35.72$ &$ 48.95$&$10.52$ &$310.43$&$178.66$&$169.70$&$7.42$ &$2.43$\\

\hline
\textbf{$C_{22}$}&$9.08$ &$4.31\times$$10^2$&$ 8.59$&$ 35.72$&$ 48.85$&$10.54$&$ 230.78$&$63.70$&$184.68$&$ 7.37$ &$4.43$\\

\hline
\textbf{$C_{31}$}&$9.23$ &$4.78\times$$10^4$&$8.55$ &$35.72$ &$ 48.50$&$10.49$& $234.75$&$185.56$&$217.34$&$7.37$&$2.44$\\

\hline
\textbf{$C_{32}$}&$9.29$&$1.09\times$$10^2$&$ 8.56$ &$ 35.72$&$ 48.94$ &$10.51$&$ 310.43$&$355.65$&$82.48$&$7.50$&$2.44$\\

\hline
\textbf{$C_{21}$}&$9.12 $&$ 2.07\times$$10^3$&$ 8.54$&$ 35.70$& $49.80$ &$10.47$&$142.15$&$279.77$&$59.23$&$7.40$& $2.43$\\

\hline
\textbf{$C_{11}$}& $9.82$& $2.29\times$$10^2$&$ 8.65$&$ 35.73$& $50.55$&$ 10.62$& $25.81$&$341.08$&$103.82$& $7.42$&$-2.49$\\
\hline
\end{tabular}
\caption{The values of $\chi_{min}^2$, $d_{FT}$, the best fit values of $\theta_{13}^\circ $, $\theta_{12}^\circ $, $\theta_{23}^\circ $, $\theta^\circ $, $\phi^\circ$, $\alpha^\circ $, $\beta^\circ$, $\Delta m^2_{21}\,{(10^{-5}\rm eV^2)}$ and $\Delta m^2_{3l}\,{(10^{-3}\rm eV^2)}$ for TM$_{2}$ mixing matrix.}
\label{tab:10}
\end{table}

In case of TM$_{1}$ mixing matrix, pattern $C_{23}$ shows very good agreement with the data with a very small $d_{FT}$ value. Although, the 
pattern $C_{31}$ also have same $\chi^2$ as pattern $C_{23}$, it, however, has a much larger $d_{FT}$ value compared to pattern $C_{23}$.
It can be concluded that for the pattern $C_{31}$, there is a strong fine tuning among the elements of the mass matrix. Similarly, 
$C_{22}$, $C_{21}$ and $C_{33}$ have larger $d_{FT}$ value compared to $C_{11}$ pattern, although they have less $\chi^2$ value than $C_{11}$.
For $C_{22}$, $C_{21}$ and $C_{33}$ also the degree of fine tuning among the mass matrix elements is very strong.
Moreover, it is very clear from Table.~\ref{tab:9} that all these patterns prefer the atmospheric mixing angle $\theta_{23}$ to be greater 
than $\pi/4$. Based on the $d_{FT}$ values, it is clear that it requires less fine tuning of the mass matrix elements for patterns $C_{23}$ 
and $C_{11}$.

For the TM$_2$ mixing matrix, the fine-tuned parameter $d_{FT}$ is small for the patterns  $C_{33}$, $C_{22}$, $C_{32}$ and $C_{11}$. Among
all these patterns $C_{32}$ has the lowest $d_{FT}$ value. However,
for patterns $C_{31}$ and $C_{21}$, $d_{FT}$ value is quite large and hence the degree of fine tuning among the elements of the mass
matrix is quite strong for these patterns. All the patterns prefer the best fit value of $\theta_{23}$ to be larger than $\pi/4$.

\section{symmetry realization}
\label{section:7}
The symmetry of one vanishing minor can be realized through type-I seesaw  mechanism~\cite{Minkowski:1977sc,Mohapatra:1979ia} along with Abelian symmetry. One vanishing minor in neutrino mass matrix can easily be obtained if one element in the Majorana matrix $M_R$ is zero 
along with diagonal Dirac mass matrix $M_D$. In order to fulfil this condition, we need three right handed charged lepton 
$l_{Rp}$ $(p=1, 2, 3)$, three right handed neutrinos $\nu_{Rp}$ and three left handed lepton doublets $D_{Lp}$. We present the symmetry 
realization of pattern V. The symmetry of this pattern can be realized through the Abelian symmetry group $(Z_{12}\times Z_2)$ that is 
discussed in Refs.~\cite{Grimus:2004hf,Dev:2010if}. 

The leptonic fields under $Z_{12}$ transform as
\begin{eqnarray}
\label{eq:50} 
&&\bar l_{R1} \rightarrow \omega \bar l_{R1}\,, \qquad\qquad
\bar \nu_{R1} \rightarrow \omega \bar \nu_{R1}\,, \qquad\qquad
D_{L1} \rightarrow \omega \bar D_{L1}\,, \nonumber \\
&&\bar l_{R2} \rightarrow \omega^2 \bar l_{R2}\,, \qquad\qquad 
\bar \nu_{R2} \rightarrow \omega^2 \bar \nu_{R2}\,, \qquad\qquad  
D_{L2} \rightarrow \omega^3 \bar D_{L2}\,, \nonumber \\ 
&&\bar l_{R3} \rightarrow \omega^5 \bar l_{R3}\,, \qquad\qquad 
\bar \nu_{R3} \rightarrow \omega^5 \bar \nu_{R3}\,, \qquad\qquad  
D_{L3} \rightarrow \omega^8 \bar D_{L3}\,,
\end{eqnarray}
where $\omega$ = ${\rm exp}(\frac{i\pi}{6})$. The bilinears $\bar l_{Rp}\,D_{Lq}$ and $\bar \nu_{Rp}\,D_{Lq}$, where $p,q = 1,2,3$, relevant 
for $(M_l)_{pq}$ and $(M_D)_{pq}$ transform as $\bar l_{Rp}\,D_{Lq} \rightarrow \Omega\,l_{Rp}\,D_{Lq}$, where  
\begin{equation}
\label{eq:51}
\Omega ={
 \begin{pmatrix}
  \omega^2& \omega^4 & \omega^9\\
 \omega^3 & \omega^5 &\omega^{10}\\
 \omega^6 & \omega^8 & \omega
\end{pmatrix}}\,
\end{equation}
and the bilinears $\bar \nu_{Rp}C \bar \nu_{Rq}^T$ relevant for $(M_R)_{pq}$ transform as $\bar \nu_{Rp}C \bar \nu_{Rq}^T \rightarrow \Lambda
\,\bar \nu_{Rp}C \bar \nu_{Rq}^T$, where
\begin{equation}
\label{eq:52}
\Lambda = {
 \begin{pmatrix}
  \omega^2& \omega^3 & \omega^6\\
 \omega^3 & \omega^4 &\omega^7\\
 \omega^6 & \omega^7 & \omega^{10}
\end{pmatrix}}\,.
\end{equation}
For each non zero element in $M_R$, we need a scalar singlet $\chi_{pq}$ and for each non zero element in $(M_l)_{pq}$ or $(M_D)_{pq}$, we 
need Higgs scalar $\phi_{pq}$ or $\tilde \phi_{pq}$, respectively. The scalar singlets get the vacuum expectation values~(vevs) at the seesaw 
scale, while Higgs doublets get vevs at the electroweak scale. Under $Z_2$ transformation, the sign of $\tilde \phi_{pq}$ and $\nu_{Rp}$ 
changes, while other multiplets remain invariant.
The diagonal charged lepton mass matrix can be obtained by introducing only three Higgs doublets namely $\phi_{11}$, $\phi_{22}$ and 
$\phi_{33}$, similarly, the diagonal Dirac neutrino mass matrix can be obtained by introducing three Higgs doublets $\tilde \phi_{11}$, 
$\tilde \phi_{22}$ and $\tilde \phi_{33}$. The non zero elements of $M_R$ can be obtained by introducing scalar fields $\chi_{11}$, 
$\chi_{12}$, $\chi_{13}$, $\chi_{22}$ and $\chi_{33}$ which under $Z_{12}$ transformation gets multiplied by $\omega^{10}$, $\omega^9$, 
$\omega^6$, $\omega^8$ and $\omega^2$, respectively. The Majorana mass matrix $M_R$ can be written as
\begin{equation}
\label{eq:53}
 M_{R}= \begin{pmatrix}
a& b & c\\
 b & d &0\\
 c & 0 & e
\end{pmatrix}\,.
\end{equation}
This provides minor zero corresponding to (3,2) element in the neutrino mass matrix. Other patterns can also be realised similarly for 
different $M_R$.

\section{conclusion}
\label{section:8}
We explore the implication of one minor zero in the neutrino mass matrix obtained using trimaximal mixing matrix. There are total six 
possible patterns and all the patterns are found to be phenomenologically compatible with the present neutrino oscillation data. The two 
unknown parameters $\theta$ and
$\phi$ of the trimaximal mixing matrix are determined by using the experimental values of the mixing angles $\theta_{12}$, $\theta_{23}$ and
$\theta_{13}$. It is found that TM$_1$ mixing matrix provides a better fit to the experimental results than TM$_2$ mixing matrix. The Jarlskog invariant measure of CP violation is non zero for all the pattern, so they are necessarily CP violating. Patterns I, II and V show normal 
mass ordering for TM$_1$ mixing matrix while these patterns show both normal and inverted mass ordering for TM$_2$ mixing matrix. Patterns III
and IV show both normal and inverted mass ordering for both TM$_1$ and TM$_2$ mixing matrix. Pattern VI predicts inverted mass ordering for 
both the mixing matrix. We predict the unknown parameters such as the absolute neutrino mass scale, the effective Majorana mass and the 
effective electron anti-neutrino mass using both TM$_1$ and TM$_2$ mixing matrix. The effective Majorana mass obtained for each pattern is 
within the reach of neutrinoless double beta decay experiment. Similarly, the value obtained for the effective electron anti-neutrino mass
may be within the reach of future Project 8 experiment. 
We also discuss the fine tuning of the elements of the mass matrix for all the patterns by introducing a new parameter $d_{FT}$. We observe 
that for the pattern $C_{23}$, the 
fine tuning among the elements of the mass matrix is small compared to other patterns.
Moreover, we also discuss the symmetry realization of pattern V using Abelian symmetry group 
$Z_{12}\times Z_2$ in the framework of type-I seesaw model which can be easily generalized to all the other patterns as well.

\appendix
\section{}
\label{app}

The coefficients in the mass ratios for the TM$_1$ mixing matrix can be expressed in terms of the two unknown parameters $\theta$ and $\phi$
as
\begin{eqnarray} 
\label{eq:56}
       && \mathcal{ A}_1=\frac{1}{6}(\frac{1}{3}\sin^2\theta+\frac{1}{2}\cos^2\theta\cos2\phi-\sqrt{\frac{2}{3}}\sin\theta \cos\theta\cos\phi),\nonumber \\
       && \mathcal{ A}_2=\frac{1}{6}(\frac{1}{2}\cos^2\theta\sin2\phi-\sqrt{\frac{2}{3}}\sin\theta \cos\theta\sin\phi),  \nonumber \\
       && \mathcal{ A}_3=(\frac{1}{3}\cos^2\theta+\frac{1}{2}\sin^2\theta\cos2\phi+\sqrt{\frac{2}{3}}\sin\theta \cos\theta\cos\phi)(\frac{1}{3}\sin^2\theta+\frac{1}{2}\cos^2\theta\cos2\phi-\sqrt{\frac{2}{3}}\sin\theta \cos\theta\cos\phi), \nonumber \\
       && \mathcal{A}_4=(\frac{1}{2}\sin^2\theta\sin2\phi+\sqrt{\frac{2}{3}}\sin\theta \cos\theta\sin\phi)(\frac{1}{2}\cos^2\theta\sin2\phi-\sqrt{\frac{2}{3}}\sin\theta \cos\theta\cos\phi),  \nonumber \\
       && \mathcal{ A}_5=(\frac{1}{2}\sin^2\theta\sin2\phi+\sqrt{\frac{2}{3}}\sin\theta \cos\theta\sin\phi)(\frac{1}{3}\sin^2\theta+\frac{1}{2}\cos^2\theta\cos2\phi-\sqrt{\frac{2}{3}}\sin\theta \cos\theta\cos\phi), \nonumber \\
       && \mathcal{ A}_6=(\frac{1}{3}\cos^2\theta+\frac{1}{2}\sin^2\theta\cos2\phi+\sqrt{\frac{2}{3}}\sin\theta \cos\theta\cos\phi)(\frac{1}{2}\cos^2\theta\sin2\phi-\sqrt{\frac{2}{3}}\sin\theta \cos\theta\sin\phi), \nonumber \\
       && \mathcal{A}_7=-\frac{1}{6}(\frac{1}{3}\cos^2\theta+\frac{1}{2}\sin^2\theta\cos2\phi+\sqrt{\frac{2}{3}}\sin\theta \cos\theta\cos\phi), \nonumber \\ 
       && \mathcal{A}_8=-\frac{1}{6}(\frac{1}{2}\sin^2\theta\sin2\phi+\sqrt{\frac{2}{3}}\sin\theta \cos\theta\sin\phi) \,.
       \end{eqnarray}

      \begin{eqnarray} 
         \label{eq:57}
       &&\mathcal{ B}_1=\frac{1}{6}(\frac{1}{3}\sin^2\theta+\frac{1}{2}\cos^2\theta\cos2\phi+\sqrt{\frac{2}{3}}\sin\theta \cos\theta\cos\phi), \nonumber \\
       &&\mathcal{ B}_2=\frac{1}{6}(\frac{1}{2}\cos^2\theta\sin2\phi+\sqrt{\frac{2}{3}}\sin\theta \cos\theta\sin\phi),  \nonumber \\
       &&\mathcal{ B}_3=(\frac{1}{3}\cos^2\theta+\frac{1}{2}\sin^2\theta\cos2\phi-\sqrt{\frac{2}{3}}\sin\theta \cos\theta\cos\phi)(\frac{1}{3}\sin^2\theta+\frac{1}{2}\cos^2\theta\cos2\phi+\sqrt{\frac{2}{3}}\sin\theta \cos\theta\cos\phi), \nonumber \\
       &&\mathcal{B}_4=(\frac{1}{2}\sin^2\theta\sin2\phi-\sqrt{\frac{2}{3}}\sin\theta \cos\theta\sin\phi)(\frac{1}{2}\cos^2\theta\sin2\phi+\sqrt{\frac{2}{3}}\sin\theta \cos\theta\cos\phi),  \nonumber \\
       &&\mathcal{ B}_5=(\frac{1}{2}\cos^2\theta\sin2\phi+\sqrt{\frac{2}{3}}\sin\theta \cos\theta\sin\phi)(\frac{1}{3}\cos^2\theta+\frac{1}{2}\sin^2\theta\cos2\phi-\sqrt{\frac{2}{3}}\sin\theta \cos\theta\cos\phi), \nonumber \\
       &&\mathcal{ B}_6=(\frac{1}{3}\sin^2\theta+\frac{1}{2}\cos^2\theta\cos2\phi+\sqrt{\frac{2}{3}}\sin\theta \cos\theta\cos\phi)(\frac{1}{2}\sin^2\theta\sin2\phi-\sqrt{\frac{2}{3}}\sin\theta \cos\theta\sin\phi), 
       \nonumber \\
       &&\mathcal{B}_7=-\frac{1}{6}(\frac{1}{3}\cos^2\theta+\frac{1}{2}\sin^2\theta\cos2\phi-\sqrt{\frac{2}{3}}\sin\theta \cos\theta\cos\phi), \nonumber \\
       &&\mathcal{B}_8=-\frac{1}{6}(\frac{1}{2}\sin^2\theta\sin2\phi-\sqrt{\frac{2}{3}}\sin\theta \cos\theta\sin\phi) \,.
       \end{eqnarray}
       
        \begin{eqnarray}
        \label{eq:58}
       &&\mathcal{C}_1=-\frac{1}{3}(\frac{1}{3}\sin\theta \cos\theta-\sqrt{\frac{1}{6}}\cos^2\theta\cos\phi), \nonumber \\
       &&\mathcal{C}_2=-\frac{1}{3\sqrt{6}}(\cos^2\theta\sin\phi),  \nonumber \\
       &&\mathcal{ C}_3=(\frac{1}{6}\sin\theta\cos^3\theta\sin^2\phi)-(\frac{1}{3}\cos^2\theta+\sqrt{\frac{1}{6}}\sin\theta \cos\theta\sin\phi)(\frac{1}{3}\sin\theta \cos\theta-\sqrt{\frac{1}{6}}\cos^2\theta\cos\phi), \nonumber \\
       &&\mathcal{C}_4=\sqrt{\frac{1}{6}}\cos^2\theta\sin\phi(\frac{1}{3}\cos^2\theta+\sqrt{\frac{1}{6}}\sin\theta \cos\theta\cos\phi)+\sqrt{\frac{1}{6}}\sin\theta \cos\theta\cos\phi(\frac{1}{3}\sin\theta \cos\theta-\sqrt{\frac{1}{6}}\cos^2\theta\cos\phi), \nonumber \\
       &&\mathcal{C}_5=\frac{1}{3}(\frac{1}{3}\cos^2\theta+\sqrt{\frac{1}{6}}\sin\theta \cos\theta\cos\phi),
        \nonumber \\
       &&\mathcal{C}_6=\frac{1}{3\sqrt{6}}\sin\theta \cos\theta\sin\phi \,.
       \end{eqnarray}
       
         \begin{eqnarray} 
         \label{eq:59}
       && \mathcal{D}_1=-\frac{1}{3}(\frac{1}{3}\sin^2\theta -\sqrt{\frac{1}{6}}\sin\theta\cos\theta\cos\phi), 
       \nonumber \\
       && \mathcal{D}_2=-\frac{1}{3\sqrt{6}}(\cos\theta\sin\phi),  \nonumber \\
       && \mathcal{D}_3=(\frac{1}{6}\sin^2\theta\cos^2\theta\sin^2\phi)+(\frac{1}{3}\cos^2\theta-\sqrt{\frac{1}{6}}\sin\theta \cos\theta\cos\phi)(\frac{1}{3}\sin^2\theta -\sqrt{\frac{1}{6}}\sin\theta\cos\theta\cos\phi), \nonumber \\
       && \mathcal{D}_4=\sqrt{\frac{1}{6}}\sin\theta\cos\theta\sin\phi(\frac{1}{3}\sin^2\theta+\sqrt{\frac{1}{6}}\sin\theta \cos\theta\cos\phi)-\sqrt{\frac{1}{6}}\sin\theta \cos\theta\sin\phi(\frac{1}{3}\cos^2\theta-\sqrt{\frac{1}{6}}\sin\theta\cos\theta\cos\phi), \nonumber \\
       && \mathcal{D}_5=\frac{1}{3}(\frac{1}{3}\cos^2\theta-\sqrt{\frac{1}{6}}\sin\theta \cos\theta\cos\phi),
        \nonumber \\
       && \mathcal{D}_6=\frac{1}{3\sqrt{6}}\sin\theta \cos\theta\sin\phi \,.
       \end{eqnarray}

       \begin{eqnarray}
       \label{eq:60}       
        &&\mathcal{E}_1=-\frac{1}{6}(\frac{1}{2}\cos^2\theta \sin2\phi-\frac{1}{2}\sin^2\theta), \nonumber \\
        &&\mathcal{E}_2=-\frac{1}{12}(\cos^2\theta\sin2\phi),  \nonumber \\
        &&\mathcal{E}_3=(\frac{1}{4}\sin^2\theta\cos^2\theta\sin^22\phi)-(\frac{1}{3}\cos^2\theta-\frac{1}{2}\sin^2\theta \cos2\phi)(\frac{1}{2}\cos^2\theta\cos2\phi-\frac{1}{3}\sin^2\theta), 
        \nonumber \\
        &&\mathcal{E}_4=\frac{1}{2}\cos^2\theta\sin2\phi(\frac{1}{3}\cos^2\theta-\frac{1}{2}\sin^2\theta\cos2\phi)+\frac{1}{2}\sin^2\theta\sin2\phi(\frac{1}{2}\cos^2\theta\cos2\phi-\frac{1}{3}\sin^2\theta), \nonumber \\
        &&\mathcal{E}_5=-\frac{1}{6}(\frac{1}{3}\cos^2\theta-\frac{1}{2}\sin^2\theta\cos2\phi), \nonumber \\ 
        &&\mathcal{E}_6=\frac{1}{12}\sin^2\theta\sin2\phi \,.
       \end{eqnarray}
     
Similarly, the coefficients in the mass ratios for the TM$_2$ mixing can be written as
    \begin{eqnarray} 
    \label{eq:61}
       &&\mathcal{\tilde A}_1=\frac{10}{36}\sin^2\theta \cos^2\theta \cos2\phi+\frac{1}{12}\sin^4\theta-\frac{1}{6\sqrt{3}}\sin^2\theta \sin2\theta \cos\phi+\frac{1}{12}\cos^4\theta(\cos^2 2\phi-\sin^2 2\phi), \nonumber \\
       &&\mathcal{\tilde A}_2=\frac{1}{6\sqrt{3}}\cos^2\theta \sin2\theta \cos3\phi-\frac{1}{12}\sin^2 2\theta \cos2\phi,  \nonumber \\
       &&\mathcal{\tilde A}_3=\frac{10}{36}\sin^2\theta \cos^2\theta \sin2\phi-\frac{1}{6\sqrt{3}}\sin^2\theta \sin2\theta \sin\phi+\frac{1}{6}\cos^4\theta \cos2\phi \sin2\phi, \nonumber \\
       &&\mathcal{\tilde A}_4=\frac{1}{6\sqrt{3}}\cos^2\theta \sin2\theta \sin3\phi-\frac{1}{12}\sin^2 2\theta \sin2\phi,  \nonumber \\
       &&\mathcal{\tilde A}_5=\frac{1}{18}\sin^2 \theta+\frac{1}{6}\cos^2\theta \cos2\phi-\frac{1}{6\sqrt{3}}\sin2\theta \cos\phi, \nonumber \\
       &&\mathcal{\tilde A}_6=\frac{1}{6}\cos^2\theta \sin2\phi-\frac{1}{6\sqrt{3}}\sin2\theta \sin\phi, \nonumber \\
       &&\mathcal{\tilde A}_7=\frac{1}{6}\sin^2 \theta+\frac{1}{18}\cos^2\theta \cos2\phi+\frac{1}{6\sqrt{3}}\sin2\theta \cos\phi,
        \nonumber \\
       &&\mathcal{\tilde A}_8=\frac{1}{18}\cos^2\theta \sin2\phi+\frac{1}{6\sqrt{3}}\sin2\theta \sin\phi \,.
       \end{eqnarray}
       
          \begin{eqnarray}
          \label{eq:62}
       &&\mathcal{\tilde B}_1=\frac{10}{36}\sin^2\theta \cos^2\theta \cos2\phi+\frac{1}{12}\sin^4\theta+\frac{1}{6\sqrt{3}}\sin^2\theta \sin2\theta \cos\phi+\frac{1}{12}\cos^4\theta(\cos^2 2\phi-\sin^2 2\phi), \nonumber \\
       &&\mathcal{\tilde B}_2=-\frac{1}{6\sqrt{3}}\cos^2\theta \sin2\theta \cos3\phi-\frac{1}{12}\sin^2 2\theta \cos2\phi,  \nonumber \\
       &&\mathcal{\tilde B}_3=\frac{10}{36}\sin^2\theta \cos^2\theta \sin2\phi+\frac{1}{6\sqrt{3}}\sin^2\theta \sin2\theta \sin\phi+\frac{1}{6}\cos^4\theta \cos2\phi \sin2\phi, \nonumber \\
       &&\mathcal{\tilde B}_4=-\frac{1}{6\sqrt{3}}\cos^2\theta \sin2\theta \sin3\phi-\frac{1}{12}\sin^2 2\theta \sin2\phi,  \nonumber \\
       &&\mathcal{\tilde B}_5=\frac{1}{18}\sin^2 \theta+\frac{1}{6}\cos^2\theta \cos2\phi+\frac{1}{6\sqrt{3}}\sin2\theta \sin\phi, \nonumber \\
       &&\mathcal{\tilde B}_6=\frac{1}{6}\cos^2\theta\sin2\phi+\frac{1}{6\sqrt{3}}\sin2\theta \sin\phi, \nonumber \\
       &&\mathcal{\tilde B}_7=\frac{1}{6}\sin^2 \theta+\frac{1}{18}\cos^2\theta \cos2\phi-\frac{1}{6\sqrt{3}}\sin2\theta \cos\phi,
       \nonumber \\
       &&\mathcal{\tilde B}_8=\frac{1}{18}\cos^2\theta \sin2\phi-\frac{1}{6\sqrt{3}}\sin2\theta \sin\phi \,.
       \end{eqnarray}
       
     \begin{eqnarray} 
     \label{eq:63}
       &&\mathcal{\tilde C}_1=\frac{1}{9}\sin^2\theta \cos^2\theta \cos2\phi-\frac{1}{6\sqrt{3}}\cos^2\theta \sin2\theta \cos3\phi+\frac{1}{6\sqrt{3}}\sin^2\theta \sin2\theta \cos\phi-\frac{1}{12}\sin^22\theta \cos2\phi, \nonumber \\
       &&\mathcal{\tilde C}_2=\frac{1}{9}\sin^2\theta \cos^2\theta \sin2\phi-\frac{1}{6\sqrt{3}}\cos^2\theta \sin2\theta \sin3\phi+\frac{1}{6\sqrt{3}}\sin^2\theta \sin2\theta \sin\phi-\frac{1}{12}\sin^2\theta \sin2\phi,  \nonumber \\
       &&\mathcal{\tilde C}_3=\frac{1}{6\sqrt{3}}\sin2\theta \cos\phi - \frac{1}{9}\sin^2\theta,  \nonumber \\
       &&\mathcal{\tilde C}_4=\frac{1}{6\sqrt{3}}\sin2\theta \sin\phi,  \nonumber \\
       &&\mathcal{\tilde C}_5=\frac{1}{9}\cos^2 \theta \cos2\phi+\frac{1}{6\sqrt{3}}\sin2\theta \cos\phi, \nonumber \\ 
       &&\mathcal{\tilde C}_6=\frac{1}{9}\cos^2\theta \sin2\phi+\frac{1}{6\sqrt{3}}\sin2\theta \sin\phi \,.
    \end{eqnarray}

    \begin{eqnarray}
    \label{eq:64}
       &&\mathcal{\tilde D}_1=\frac{1}{9}\sin^2\theta \cos^2\theta \cos2\phi-\frac{1}{6\sqrt{3}}\sin^2\theta \sin2\theta \cos3\phi+\frac{1}{6\sqrt{3}}\cos^2\theta \sin2\theta \cos\phi-\frac{1}{12}\sin^22\theta \cos2\phi, \nonumber \\
       &&\mathcal{\tilde D}_2=\frac{1}{9}\sin^2\theta \cos^2\theta \sin2\phi-\frac{1}{6\sqrt{3}}\sin^2\theta \sin2\theta \sin\phi+\frac{1}{6\sqrt{3}}\cos^2\theta \sin2\theta \sin3\phi-\frac{1}{12}\sin^2\theta \sin2\phi,  \nonumber \\
       &&\mathcal{\tilde D}_3=\frac{1}{6\sqrt{3}}\sin2\theta \cos\phi + \frac{1}{9}\sin^2\theta, \nonumber \\
       &&\mathcal{\tilde D}_4=\frac{1}{6\sqrt{3}}\sin2\theta \sin\phi,  \nonumber \\
       &&\mathcal{\tilde D}_5=\frac{1}{9}\cos^2 \theta\cos2\phi-\frac{1}{6\sqrt{3}}\sin2\theta \cos\phi, \nonumber \\ 
       &&\mathcal{\tilde D}_6=\frac{1}{6\sqrt{3}}\sin2\theta \sin\phi-\frac{1}{9}\cos^2\theta \sin2\phi \,.
    \end{eqnarray}
    
     \begin{eqnarray} 
     \label{eq:65}
       &&\mathcal{\tilde E}_1=\frac{10}{36}\sin^2\theta \cos^2\theta \cos2\phi-\frac{1}{12}\cos^4\theta \cos^2 2\phi-\frac{1}{12}\sin^4\theta +\frac{1}{12}\cos^4\theta \sin^2 2\phi, \nonumber \\
       &&\mathcal{\tilde E}_2=\frac{5}{18}\sin^2\theta \cos^2\theta \sin2\phi-\frac{1}{6}\cos^4\theta \sin2\phi \cos2\phi, \nonumber \\
       &&\mathcal{\tilde E}_3=\frac{1}{6}\cos^2\theta \cos2\phi - \frac{1}{18}\sin^2\theta,  \nonumber \\
       &&\mathcal{\tilde E}_4=\frac{1}{6}\cos^2\theta \sin2\phi,  \nonumber \\
       &&\mathcal{\tilde E}_5=\frac{1}{6}\sin^2 \theta - \frac{1}{18}\cos^2\theta \cos2\phi,
        \nonumber \\
       &&\mathcal{\tilde E}_6=\frac{1}{6}\cos^2\theta \sin2\phi\,.
      \end{eqnarray}


\bigskip    
\begin{thebibliography}{99}

\bibitem{Super-Kamiokande:1998kpq}
Y.~Fukuda \textit{et al.} [Super-Kamiokande],
Phys. Rev. Lett. \textbf{81}, 1562-1567 (1998)
doi:10.1103/PhysRevLett.81.1562
[arXiv:hep-ex/9807003 [hep-ex]].\\

\bibitem{Project8:2022wqh}A.~A.~Esfahani \textit{et al.} [Project 8],[arXiv:2203.07349 [nucl-ex]].\\

\bibitem{Zhang:2020mox}
M.~Zhang, J.~F.~Zhang and X.~Zhang,
Commun. Theor. Phys. \textbf{72}, no.12, 125402 (2020)
[arXiv:2005.04647 [astro-ph.CO]].\\

\bibitem{GERDA:2013vls}
M.~Agostini \textit{et al.} [GERDA],
Phys. Rev. Lett. \textbf{111}, no.12, 122503 (2013)
doi:10.1103/PhysRevLett.111.122503
[arXiv:1307.4720 [nucl-ex]].\\

\bibitem{CUORE:2015hsf}
K.~Alfonso \textit{et al.} [CUORE],
Phys. Rev. Lett. \textbf{115}, no.10, 102502 (2015)
doi:10.1103/PhysRevLett.115.102502
[arXiv:1504.02454 [nucl-ex]].\\

\bibitem{EXO-200:2014ofj}
J.~B.~Albert \textit{et al.} [EXO-200],
Nature \textbf{510}, 229-234 (2014)
doi:10.1038/nature13432
[arXiv:1402.6956 [nucl-ex]].\\

\bibitem{KamLAND-Zen:2012mmx}
A.~Gando \textit{et al.} [KamLAND-Zen],
Phys. Rev. Lett. \textbf{110}, no.6, 062502 (2013)
doi:10.1103/PhysRevLett.110.062502
[arXiv:1211.3863 [hep-ex]].\\

\bibitem{Lashin:2011dn}
E.~I.~Lashin and N.~Chamoun,
Phys. Rev. D \textbf{85}, 113011 (2012)
doi:10.1103/PhysRevD.85.113011
[arXiv:1108.4010 [hep-ph]].\\

\bibitem{Singh:2018tqu}
M.~Singh,
Adv. High Energy Phys. \textbf{2018}, 2863184 (2018)
doi:10.1155/2018/2863184
[arXiv:1803.10735 [hep-ph]].\\

\bibitem{Gautam:2018izb}
R.~R.~Gautam,
Phys. Rev. D \textbf{97}, no.5, 055022 (2018)
doi:10.1103/PhysRevD.97.055022
[arXiv:1802.00425 [hep-ph]].\\

\bibitem{Frampton:2002yf}
P.~H.~Frampton, S.~L.~Glashow and D.~Marfatia,
Phys. Lett. B \textbf{536}, 79-82 (2002)
doi:10.1016/S0370-2693(02)01817-8
[arXiv:hep-ph/0201008 [hep-ph]].\\


\bibitem{Xing:2002ta}
Z.~z.~Xing,
Phys. Lett. B \textbf{530}, 159-166 (2002)
doi:10.1016/S0370-2693(02)01354-0
[arXiv:hep-ph/0201151 [hep-ph]].\\

\bibitem{Xing:2002ap}
Z.~z.~Xing,
Phys. Lett. B \textbf{539}, 85-90 (2002)
doi:10.1016/S0370-2693(02)02062-2
[arXiv:hep-ph/0205032 [hep-ph]].\\

\bibitem{Lavoura:2004tu}
L.~Lavoura,
Phys. Lett. B \textbf{609}, 317-322 (2005)
doi:10.1016/j.physletb.2005.01.047
[arXiv:hep-ph/0411232 [hep-ph]].\\

\bibitem{Dev:2006qe}
S.~Dev, S.~Kumar, S.~Verma and S.~Gupta,
Phys. Rev. D \textbf{76}, 013002 (2007)
doi:10.1103/PhysRevD.76.013002
[arXiv:hep-ph/0612102 [hep-ph]].\\


\bibitem{Kumar:2011vf}
S.~Kumar,
Phys. Rev. D \textbf{84}, 077301 (2011)
doi:10.1103/PhysRevD.84.077301
[arXiv:1108.2137 [hep-ph]].\\

\bibitem{Fritzsch:2011qv}
H.~Fritzsch, Z.~z.~Xing and S.~Zhou,
JHEP \textbf{09}, 083 (2011)
doi:10.1007/JHEP09(2011)083
[arXiv:1108.4534 [hep-ph]].\\

\bibitem{Ludl:2011vv}
P.~O.~Ludl, S.~Morisi and E.~Peinado,
Nucl. Phys. B \textbf{857}, 411-423 (2012)
doi:10.1016/j.nuclphysb.2011.12.017
[arXiv:1109.3393 [hep-ph]].\\

\bibitem{Meloni:2012sx}
D.~Meloni and G.~Blankenburg,
Nucl. Phys. B \textbf{867}, 749-762 (2013)
doi:10.1016/j.nuclphysb.2012.10.011
[arXiv:1204.2706 [hep-ph]].\\

\bibitem{Grimus:2012zm}
W.~Grimus and P.~O.~Ludl,
J. Phys. G \textbf{40}, 055003 (2013)
doi:10.1088/0954-3899/40/5/055003
[arXiv:1208.4515 [hep-ph]].\\Dev:2015lya

\bibitem{Dev:2014dla}
S.~Dev, R.~R.~Gautam, L.~Singh and M.~Gupta,
Phys. Rev. D \textbf{90}, no.1, 013021 (2014)
doi:10.1103/PhysRevD.90.013021
[arXiv:1405.0566 [hep-ph]].\\


\bibitem{Gautam:2016qyw}
R.~R.~Gautam and S.~Kumar,
Phys. Rev. D \textbf{94}, no.3, 036004 (2016)
[erratum: Phys. Rev. D \textbf{100}, no.3, 039902 (2019)]
doi:10.1103/PhysRevD.94.036004
[arXiv:1607.08328 [hep-ph]].\\

\bibitem{Channey:2018cfj}
K.~S.~Channey and S.~Kumar,
J. Phys. G \textbf{46}, no.1, 015001 (2019)
doi:10.1088/1361-6471/aaf55e
[arXiv:1812.10268 [hep-ph]].\\

\bibitem{Singh:2019baq}
M.~Singh,
EPL \textbf{129}, no.1, 1 (2020)
doi:10.1209/0295-5075/129/11002
[arXiv:1909.01552 [hep-ph]].\\

\bibitem{Lashin:2007dm}
E.~I.~Lashin and N.~Chamoun,
Phys. Rev. D \textbf{78}, 073002 (2008)
doi:10.1103/PhysRevD.78.073002
[arXiv:0708.2423 [hep-ph]].\\

\bibitem{Lashin:2009yd}
E.~I.~Lashin and N.~Chamoun,
Phys. Rev. D \textbf{80}, 093004 (2009)
doi:10.1103/PhysRevD.80.093004
[arXiv:0909.2669 [hep-ph]].\\

\bibitem{Dev:2010if}
S.~Dev, S.~Verma, S.~Gupta and R.~R.~Gautam,0
Phys. Rev. D \textbf{81}, 053010 (2010)
doi:10.1103/PhysRevD.81.053010
[arXiv:1003.1006 [hep-ph]].\\


\bibitem{Dev:2010pf}
S.~Dev, S.~Gupta and R.~R.~Gautam,
Mod. Phys. Lett. A \textbf{26}, 501-514 (2011)
doi:10.1142/S0217732311034906
[arXiv:1011.5587 [hep-ph]].\\

\bibitem{Tavartkiladze:2022pzf}
Z.~Tavartkiladze,
Phys. Rev. D \textbf{106}, no.11, 115002 (2022)
doi:10.1103/PhysRevD.106.115002
[arXiv:2209.14404 [hep-ph]].\\

\bibitem{Araki:2012ip}
T.~Araki, J.~Heeck and J.~Kubo,
JHEP \textbf{07}, 083 (2012)
doi:10.1007/JHEP07(2012)083
[arXiv:1203.4951 [hep-ph]].\\

\bibitem{Liao:2013saa}
J.~Liao, D.~Marfatia and K.~Whisnant,
JHEP \textbf{09}, 013 (2014)
doi:10.1007/JHEP09(2014)013
[arXiv:1311.2639 [hep-ph]].\\


\bibitem{Dev:2013xca}
S.~Dev, R.~R.~Gautam and L.~Singh,
Phys. Rev. D \textbf{87}, 073011 (2013)
doi:10.1103/PhysRevD.87.073011
[arXiv:1303.3092 [hep-ph]].\\

\bibitem{Wang:2013woa}
W.~Wang,
Eur. Phys. J. C \textbf{73}, 2551 (2013)
doi:10.1140/epjc/s10052-013-2551-2
[arXiv:1306.3556 [hep-ph]].\\


\bibitem{Whisnant:2015ovx}
K.~Whisnant, J.~Liao and D.~Marfatia,
AIP Conf. Proc. \textbf{1604}, no.1, 273-278 (2015)
doi:10.1063/1.4883441\\

\bibitem{Dev:2015lya}
S.~Dev, L.~Singh and D.~Raj,
Eur. Phys. J. C \textbf{75}, no.8, 394 (2015)
doi:10.1140/epjc/s10052-015-3569-4
[arXiv:1506.04951 [hep-ph]].\\

\bibitem{Wang:2016tkm}
W.~Wang, S.~Y.~Guo and Z.~G.~Wang,
Mod. Phys. Lett. A \textbf{31}, no.13, 1650080 (2016)
doi:10.1142/S0217732316500802\\


\bibitem{Goswami:2008uv}
S.~Goswami, S.~Khan and A.~Watanabe,
Phys. Lett. B \textbf{693}, 249-254 (2010)
doi:10.1016/j.physletb.2010.08.033
[arXiv:0811.4744 [hep-ph]].\\



\bibitem{Dev:2009he}
S.~Dev, S.~Verma and S.~Gupta,
Phys. Lett. B \textbf{687}, 53-60 (2010)
doi:10.1016/j.physletb.2010.02.055
[arXiv:0909.3182 [hep-ph]].\\

\bibitem{Dev:2010pe}
S.~Dev, S.~Gupta and R.~R.~Gautam,
Phys. Rev. D \textbf{82}, 073015 (2010)
doi:10.1103/PhysRevD.82.073015
[arXiv:1009.5501 [hep-ph]].\\


\bibitem{Liu:2013oxa}
J.~Y.~Liu and S.~Zhou,
Phys. Rev. D \textbf{87}, no.9, 093010 (2013)
doi:10.1103/PhysRevD.87.093010
[arXiv:1304.2334 [hep-ph]].\\

\bibitem{Dev:2013nua}
S.~Dev, R.~R.~Gautam and L.~Singh,
Phys. Rev. D \textbf{88}, 033008 (2013)
doi:10.1103/PhysRevD.88.033008
[arXiv:1306.4281 [hep-ph]].\\


\bibitem{Harrison:2002er}
P.~F.~Harrison, D.~H.~Perkins and W.~G.~Scott,
Phys. Lett. B \textbf{530}, 167 (2002)
doi:10.1016/S0370-2693(02)01336-9
[arXiv:hep-ph/0202074 [hep-ph]].\\


\bibitem{Harrison:2002kp}
P.~F.~Harrison and W.~G.~Scott,
Phys. Lett. B \textbf{535}, 163-169 (2002)
doi:10.1016/S0370-2693(02)01753-7
[arXiv:hep-ph/0203209 [hep-ph]].\\


\bibitem{Xing:2002sw}
Z.~z.~Xing,
Phys. Lett. B \textbf{533}, 85-93 (2002)
doi:10.1016/S0370-2693(02)01649-0
[arXiv:hep-ph/0204049 [hep-ph]].\\


\bibitem{Harrison:2003aw}
P.~F.~Harrison and W.~G.~Scott,
Phys. Lett. B \textbf{557}, 76 (2003)
doi:10.1016/S0370-2693(03)00183-7
[arXiv:hep-ph/0302025 [hep-ph]].\\


\bibitem{T2K:2011ypd}
K.~Abe \textit{et al.} [T2K],
Phys. Rev. Lett. \textbf{107}, 041801 (2011)
doi:10.1103/PhysRevLett.107.041801
[arXiv:1106.2822 [hep-ex]].\\



\bibitem{MINOS:2011amj}
P.~Adamson \textit{et al.} [MINOS],
Phys. Rev. Lett. \textbf{107}, 181802 (2011)
doi:10.1103/PhysRevLett.107.181802
[arXiv:1108.0015 [hep-ex]].\\



\bibitem{DoubleChooz:2011ymz}
Y.~Abe \textit{et al.} [Double Chooz],
Phys. Rev. Lett. \textbf{108}, 131801 (2012)
doi:10.1103/PhysRevLett.108.131801
[arXiv:1112.6353 [hep-ex]].\\


\bibitem{DayaBay:2012fng}
F.~P.~An \textit{et al.} [Daya Bay],
Phys. Rev. Lett. \textbf{108}, 171803 (2012)
doi:10.1103/PhysRevLett.108.171803
[arXiv:1203.1669 [hep-ex]].\\


\bibitem{RENO:2012mkc}
J.~K.~Ahn \textit{et al.} [RENO],
Phys. Rev. Lett. \textbf{108}, 191802 (2012)
doi:10.1103/PhysRevLett.108.191802
[arXiv:1204.0626 [hep-ex]].\\

\bibitem{Kumar:2010qz}
S.~Kumar,
Phys. Rev. D \textbf{82}, 013010 (2010)
[erratum: Phys. Rev. D \textbf{85}, 079904 (2012)]
doi:10.1103/PhysRevD.82.013010
[arXiv:1007.0808 [hep-ph]].\\

\bibitem{He:2011gb}
X.~G.~He and A.~Zee,
Phys. Rev. D \textbf{84}, 053004 (2011)
doi:10.1103/PhysRevD.84.053004
[arXiv:1106.4359 [hep-ph]].\\

\bibitem{Grimus:2008tt}
W.~Grimus and L.~Lavoura,
JHEP \textbf{09}, 106 (2008)
doi:10.1088/1126-6708/2008/09/106
[arXiv:0809.0226 [hep-ph]].\\



\bibitem{Jarlskog:1985ht}
C.~Jarlskog,
Phys. Rev. Lett. \textbf{55}, 1039 (1985)
doi:10.1103/PhysRevLett.55.1039\\


\bibitem{Esteban:2020cvm}
I.~Esteban, M.~C.~Gonzalez-Garcia, M.~Maltoni, T.~Schwetz and A.~Zhou,
JHEP \textbf{09}, 178 (2020)
doi:10.1007/JHEP09(2020)178
[arXiv:2007.14792 [hep-ph]].NuFIT 5.0(2020), http://www.nu-fit.org.\\


\bibitem{Altarelli:2010at}
G.~Altarelli and G.~Blankenburg,
JHEP \textbf{03}, 133 (2011)
doi:10.1007/JHEP03(2011)133
[arXiv:1012.2697 [hep-ph]].

\bibitem{Minkowski:1977sc}
P.~Minkowski,
Phys. Lett. B \textbf{67}, 421-428 (1977)
doi:10.1016/0370-2693(77)90435-X.\\

\bibitem{Mohapatra:1979ia}
R.~N.~Mohapatra and G.~Senjanovic,
Phys. Rev. Lett. \textbf{44}, 912 (1980)
doi:10.1103/PhysRevLett.44.912.\\


\bibitem{Grimus:2004hf}
W.~Grimus, A.~S.~Joshipura, L.~Lavoura and M.~Tanimoto,
Eur. Phys. J. C \textbf{36}, 227-232 (2004)
doi:10.1140/epjc/s2004-01896-y
[arXiv:hep-ph/0405016 [hep-ph]].\\



\end{thebibliography}
\end{document}